\documentclass[aps,prb,english,reprint,preprintnumbers,superscriptaddress,amsmath,amssymb,bibnotes,longbibliography]{revtex4-2}
\usepackage{graphicx}
\usepackage{dcolumn}
\usepackage{epstopdf}
\usepackage{bm}
\usepackage{array}
\usepackage[pdfstartview=FitH, CJKbookmarks=true, bookmarksnumbered=true, bookmarksopen=true, colorlinks, linkcolor=blue, anchorcolor=blue, citecolor=blue,urlcolor=blue,breaklinks]{hyperref}
\usepackage[mathlines]{lineno}
\usepackage{float}
\usepackage{makecell}

\usepackage{amsmath}
\usepackage{array}
\newcolumntype{L}[1]{>{\raggedright\arraybackslash}m{#1}}
\newcolumntype{C}[1]{>{\centering\arraybackslash}m{#1}}

\makeatletter
\@ifundefined{textcolor}{}
{%
 \definecolor{BLACK}{gray}{0}
 \definecolor{WHITE}{gray}{1}
 \definecolor{RED}{rgb}{1,0,0}
 \definecolor{GREEN}{rgb}{0,1,0}
 \definecolor{BLUE}{rgb}{0,0,1}
 \definecolor{CYAN}{cmyk}{1,0,0,0}
 \definecolor{MAGENTA}{cmyk}{0,1,0,0}
 \definecolor{YELLOW}{cmyk}{0,0,1,0}
}

\usepackage{color}

\newcommand{\beq}{\begin{eqnarray}}
\newcommand{\eeq}{\end{eqnarray}}
\newcommand{\ys}[1]{\textcolor{black}{#1}}
\newcommand{\KFAT}{KFAT}
\makeatother
\usepackage{babel}

\begin{document}
\raggedbottom

\title{Twinning of domains and spin anisotropy in K$_5$Fe$_4$Ag$_6$Te$_{10}$}

\author{Jiayu Guo}
\affiliation{Center for Correlated Matter and School of Physics, Zhejiang University, Hangzhou 310058, China}

\author{Hengyang Zhong}
\affiliation{Center for Advanced Quantum Studies and Department of Physics, Beijing Normal University, Beijing 100875, China}

\author{Dongsheng Yuan}
\affiliation{Materials Science Division, Lawrence Berkeley National Lab, Berkeley, California 94720, USA}
\affiliation{\ys{State Key Laboratory of Functional Crystals and Devices, Shanghai Institute of Ceramics, Chinese Academy of Sciences, Shanghai 201899, China}}

\author{Xuejuan Gui}
\affiliation{Laboratory for Neutron Scattering, School of Physics, Renmin University of China, Beijing 100872, China}
\affiliation{Key Laboratory of Quantum State Construction and Manipulation (Ministry of Education), Renmin University of China, Beijing, 100872, China}

\author{Youzhe Chen}
\affiliation{Department of Physics, University of California, Berkeley, California 94720, USA}
\affiliation{Materials Science Division, Lawrence Berkeley National Lab, Berkeley, California 94720, USA}

\author{Nathan Giles-Donovan}
\affiliation{Department of Physics, University of California, Berkeley, California 94720, USA}
\affiliation{Materials Science Division, Lawrence Berkeley National Lab, Berkeley, California 94720, USA}
\author{\ys{Naomi Kawamura}}
\affiliation{\ys{Japan Synchrotron Radiation Research Institute, SPring-8, 1-1-1 Kouto, Sayo, Hyogo 679-5198, Japan}}

\author{Masaaki Matsuda}
\affiliation{Neutron Scattering Division, Oak Ridge National Laboratory, Oak Ridge, Tennessee 37831, USA}

\author{Yaohua Liu}
\affiliation{Neutron Scattering Division, Oak Ridge National Laboratory, Oak Ridge, Tennessee 37831, USA}
\affiliation{Second Target Station, Oak Ridge National Laboratory, Oak Ridge, Tennessee 37831, USA}

\author{Feng Ye}
\affiliation{Neutron Scattering Division, Oak Ridge National Laboratory, Oak Ridge, Tennessee 37831, USA}


\author{Rongyan Chen}
\affiliation{Center for Advanced Quantum Studies and Department of Physics, Beijing Normal University, Beijing 100875, China}

\author{Robert J. Birgeneau}
\affiliation{Department of Physics, University of California, Berkeley, California 94720, USA}
\affiliation{Materials Science Division, Lawrence Berkeley National Lab, Berkeley, California 94720, USA}

\author{Xingye Lu}
\affiliation{Center for Advanced Quantum Studies and Department of Physics, Beijing Normal University, Beijing 100875, China}

\author{Jincheng Wang}
\email{jcwang\_phys@ruc.edu.cn}
\affiliation{Laboratory for Neutron Scattering, School of Physics, Renmin University of China, Beijing 100872, China}
\affiliation{Key Laboratory of Quantum State Construction and Manipulation (Ministry of Education), Renmin University of China, Beijing, 100872, China}
\affiliation{PSI Center for Neutron and Muon Sciences, 5232 Villigen PSI, Switzerland}
\affiliation{Laboratory for Quantum Magnetism, Institute of Physics, École Polytechnique Fédérale de Lausanne (EPFL), 1015 Lausanne, Switzerland}

\author{Yu Song}
\email{yusong\_phys@zju.edu.cn}
\affiliation{Center for Correlated Matter and School of Physics, Zhejiang University, Hangzhou 310058, China}

\selectlanguage{english}%

\begin{abstract}
The Fe-based superconductors are derived from metallic parent compounds with nematic and stripe magnetic orders, which lead to two types of magnetic domains. Recently it was found that K$_5$Fe$_4$Ag$_6$Te$_{10}$ ({\KFAT}), an Fe-based semiconductor, exhibits similar nematic and stripe magnetic orders, and is thus an analogue to the Fe-based superconductors in the limit of localized electrons. In this work, the superstructure and magnetic domains of {\KFAT} are elucidated by fully mapping the reciprocal space using time-of-flight single crystal neutron diffraction. In {\KFAT}, Fe and Ag atoms order to form a $\sqrt{5}\times\sqrt{5}$ superstructure containing $2\times2$ Fe blocks, which leads to two superstructure domains with identical main Bragg peaks but distinct superstructure peaks. Below $T_{\rm N}\approx35$~K, magnetic and nematic orders break in-plane rotational symmetry of the tetragonal $\sqrt{5}\times\sqrt{5}$ superstructure, and further give rise to two magnetic domains. \ys{These four equally populated} domains account for the complex scattering pattern observed in our time-of-flight elastic neutron scattering measurements. Using polarized neutron scattering, we demonstrate a prominent spin anisotropy with an easy-plane spanned by the $c$-axis and the intra-block antiferromagnetic Fe-Fe bond direction. Such an anisotropy at ${\bf q}\neq0$ persists well above $T_{\rm N}$, \ys{accounts} for the in-plane ${\bf q}=0$ magnetic anisotropy observed in uniaxial-strained {\KFAT}\ys{, and offers an indicator for discovering similar piezomagnetic effects in other materials}. 

\end{abstract}


\maketitle

\section{Introduction}

The parent compounds of Fe-based superconductors (FeSCs) are bad metals with coupled nematic and stripe antiferromagnetic (AFM) orders [${\bf q}_{\rm s}=(1/2,1/2)$ or $(1/2,-1/2)$ in the 2-Fe unit cell]. The appearance of optimal superconductivity around a putative quantum critical point of these orders suggests that the corresponding fluctuations may drive superconductivity \cite{Dai2015,si2016,Bhmer2022,Fernandes2022}. The origins of stripe magnetism and nematicity are thus important for understanding the pairing mechanism in the FeSCs, with theories that start from the itinerant \cite{Singh2008,mazin2010} and localized limits \cite{Si2008}, as well as ones that incorporate both band and localized electrons \cite{Yin2010}. 

In the scenario of stripe magnetism and nematicity originating from localized electrons, it should be possible to realize these orders in insulating analogues of the FeSCs. While stripe magnetism (characterized by magnetic ordering at ${\bf q}_{\rm s}$, or by AFM ordering along one in-plane direction and ferromagnetic along the other) were reported in CsFe$_2$Se$_3$ and BaFe$_2$S$_3$ \cite{Du2012,Takahashi2015}, K$_2$Fe$_3$Se$_4$ and Rb$_2$Fe$_3$S$_4$ \cite{Zhao2012,Wang2015}, NaFe$_{1-x}$Cu$_x$As ($x\approx0.5$) \cite{Song2016,Wang2026}, and (Fe$_{1-x}$Cu$_x$)$_{1+y}$Te \cite{Cao2024}, the crystal structures of these insulating/semiconducting compounds are twofold symmetric, and do not allow for fluctuations between Ising-nematic configurations. On the other hand, nematic fluctuations were reported in Mott insulating oxychalchogenides La$_2$O$_2$Fe$_2$O$M_2$ ($M=$~S, Se), although these compounds order at ${\bf q}=(1/2,0)$ rather than at ${\bf q}_{\rm s}$ \cite{Free2010,Freelon2021}. 

Recently, it was shown that K$_5$Fe$_4$Ag$_6$Te$_{10}$ (or KFe$_{0.8}$Ag$_{1.2}$Te$_2$, {\KFAT}), a semiconducting analogue of BaFe$_2$As$_2$ \cite{Lei2011,Ang2013}, exhibits intertwined nematic and magnetic orders \cite{Song2019} below $T_{\rm nem}\approx T_{\rm N}\approx35$~K. Fe-Ag ordering leads to $2\times2$ Fe blocks separated by Ag atoms, and gives rise to a $\sqrt{5}\times\sqrt{5}$ expansion of the tetragonal unit cell in the $ab$-plane \cite{Song2019}, similar to the superstructures observed in 245-type iron chalcogenides such as K$_2$Fe$_4$Se$_5$ \cite{Bao2011,Wang2011,Ye2011}. Below $T_{\rm nem}$, fourfold rotational symmetry is broken by the elongation of the lattice along one Fe-Fe direction, accompanied by the contraction along the other Fe-Fe direction. Below $T_{\rm N}$, the spins inside each $2\times2$ Fe block form a stripe-type configuration, displaying AFM alignment along the longer Fe-Fe direction and ferromagnetic alignment along the shorter Fe-Fe direction. Initial neutron scattering measurements suggested $T_{\rm nem}=T_{\rm N}$ \cite{Song2019}, while a more recent study combining neutron and inelastic X-ray scattering measurements evidence a first-order nematic transition that preempts a second-order magnetic transition, with $T_{\rm nem}>T_{\rm N}$ by approximately 1~K \cite{Giles-Donovan2025}. Above $T_{\rm nem}$, where static magnetic order is absent, uniaxial strain induces an in-plane anisotropy in the uniform (${\bf q}=0$) magnetic susceptibility that increases significantly upon cooling \cite{Song2019a}, similar to the divergent nematic susceptibility in the FeSCs and their metallic parent phases, observed via elastoresistivity measurements \cite{Chu2012}. These results demonstrate {\KFAT} to be an insulating/semiconducting Fe chalcogenide that realizes nematic and stripe magnetic orders resembling those in the FeSCs.

\ys{The combination of intertwined nematic and magnetic orders with a $\sqrt{5}\times\sqrt{5}$ superstructure in {\KFAT} suggests complex twinning and domains, an understanding of which is essential for the interpretation of experimental results on this system. Furthermore, the magnetic structure of {\KFAT} suggests a hierarchy of spin anisotropy similar to that in the FeSCs \cite{Song2019}, with the associated spin-anisotropic fluctuations directly responsible for the strain-induced spin-nematic state above $T_{\rm nem}$ (characterized by a ${\bf q}=0$ magnetic susceptibility anisotropy) \cite{Song2019a}. Therefore, these spin anisotropic fluctuations are not only integral for understanding the origin of nematicity in {\KFAT}, but also hold broader implications for the ubiquitous nematicity in the FeSCs.}

In this work, we use synchrotron X-ray diffraction and time-of-flight neutron diffraction to \ys{elucidate the twinning of superstructure and magnetic domains in {\KFAT} via full mapping of the reciprocal space. We show that the rather complex scattering patterns arise from twinning of the same twofold symmetric single-${\bf q}$ incommensurate magnetic structure, with equally populated domains yielding an apparent fourfold symmetry in the diffraction data. By refining the magnetic structure against the neutron diffraction data, we find that a noncollinear intra-block spin arrangement provides the best fit, suggesting a more complex magnetic ground state than previously reported. The spin anisotropy in {\KFAT} is probed directly at the magnetic ordering vector using polarized neutron scattering, and spin-anisotropic fluctuations are shown to persist well above $T_{\rm nem}$ and are directly responsible for the strain-induced in-plane anisotropy of the uniform magnetic susceptibility. Optical reflectivity measurements reveal a $\sim1.09$~eV direct charge gap, confirming the bulk semiconducting/insulating nature of {\KFAT}.}

\begin{figure}
    \includegraphics[width=1\columnwidth]{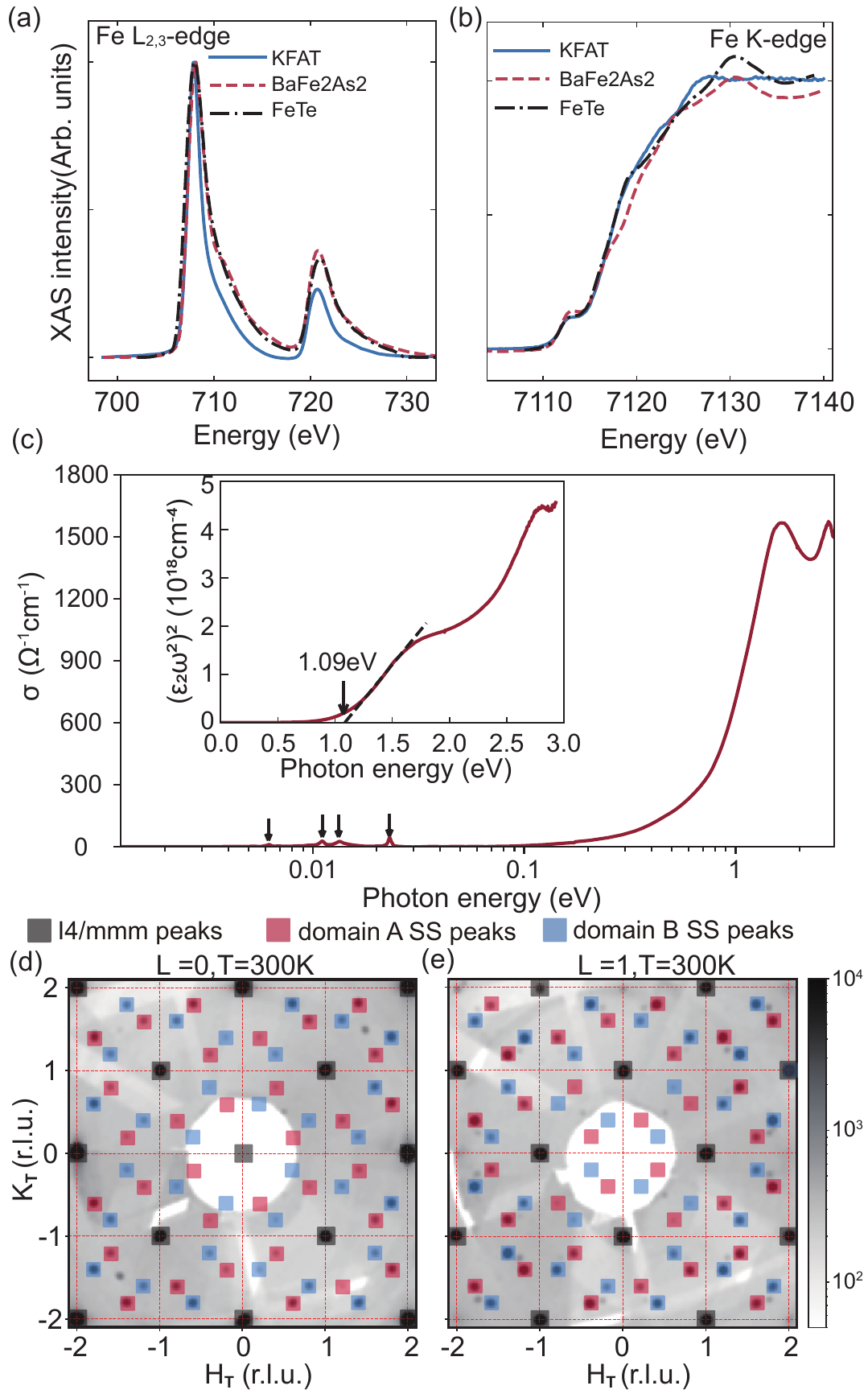}
    \protect\caption{\ys{
    (a) Comparison of Fe $L_{2,3}$-edge X-ray absorption spectra of BaFe$_2$As$_2$, {\KFAT} and FeTe~\cite{saini2011}.
    (b) Comparison of Fe $K$-edge X-ray absorption spectra of BaFe$_2$As$_2$~\cite{Bittar2011CoSubstitution}, FeTe~\cite{joseph2010} and {\KFAT}.
    (c) Optical conductivity $\sigma_1(\omega)$ of {\KFAT} obtained from room-temperature reflectivity measurements. The arrows indicate the low-energy phonon modes. The inset shows the Tauc analysis of the direct band gap of {\KFAT} at room temperature. The red solid curve represents $(\varepsilon_2\omega^2)^2$, with a linear region near the absorption edge indicated by the dashed line.
    Single-crystal XRD maps measured at 300~K in the (d) $[HK0]_{\rm T}$ and (e) $[HK1]_{\rm T}$ planes. Expected Bragg peak positions are marked by square symbols.
    }}
    \label{basic}

\end{figure}
\begin{figure*}
	\includegraphics[width=2\columnwidth]{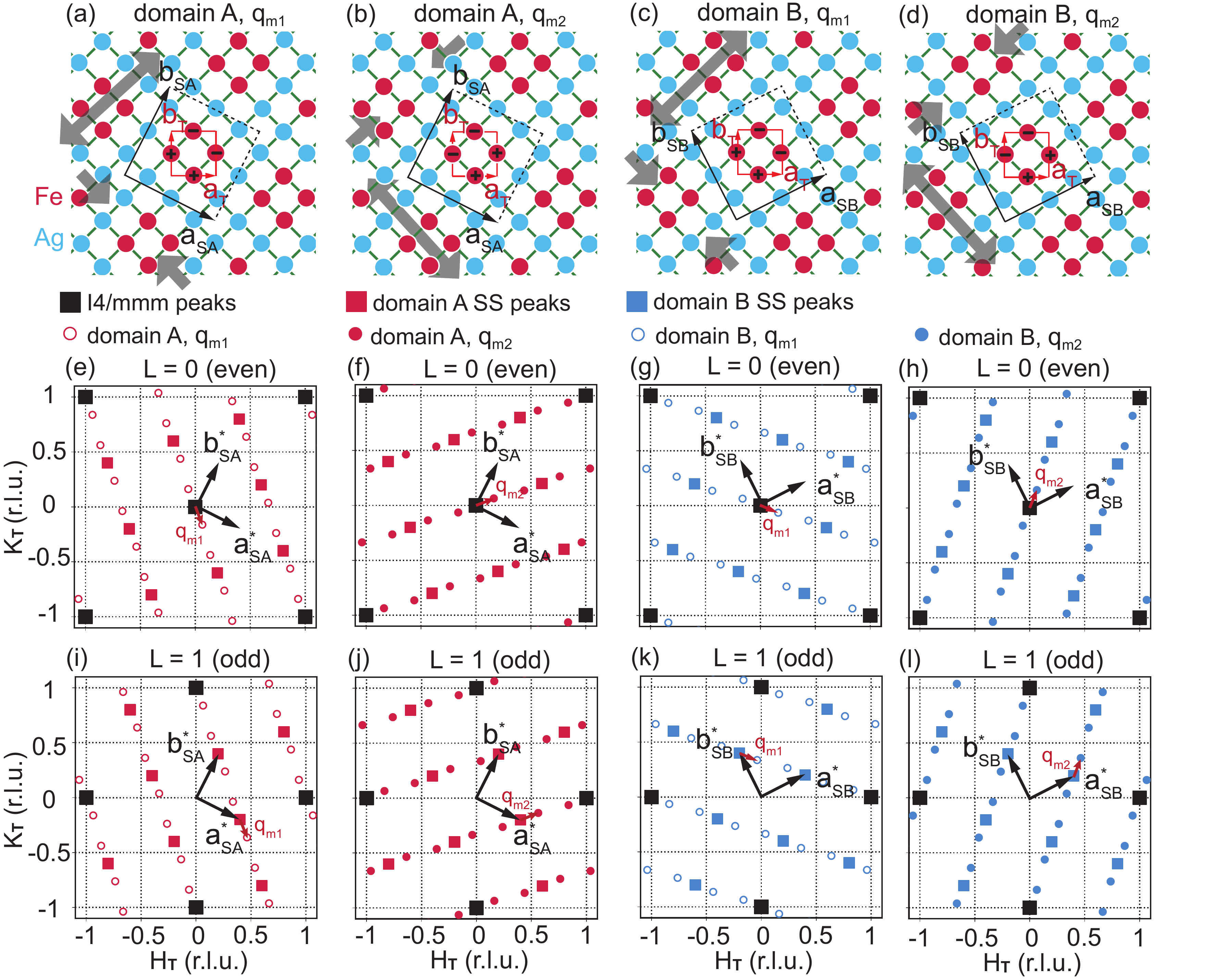} \protect\caption{(a)–(d) Schematic Fe–Ag planes in {\KFAT}. The red solid boxes denote the $I4/mmm$ unit cells, while the black boxes indicate the $\sqrt{5} \times \sqrt{5}$ superstructure. The `+' and `–'  symbols represent AFM magnetic moments. The two superstructure domains (A and B) each gives rise to two magnetic domains respectively modulated by $\mathbf{q}_{\rm m1}$ and $\mathbf{q}_{\rm m2}$. The gray arrows indicate that the lattice is elongated along the AFM Fe-Fe direction and compressed along the FM Fe-Fe direction. (e)–(l) Schematic scattering patterns of {\KFAT}. (e)–(h) Even-$L$ $HK$-planes for domains in (a)-(d). (i)–(l) Odd-$L$ $HK$-planes for domains in (a)-(d). Black squares are main Bragg peaks of the $I4/mmm$ lattice. Solid red and blue squares are superstructure peaks of domain A (domain A SS peaks) and domain B (domain B SS peaks), respectively. Open and closed red circles are magnetic peaks of domain A modulated by $\mathbf{q}_{\rm m1}$ and $\mathbf{q}_{\rm m2}$, respectively. Similarly, the $\mathbf{q}_{\rm m1}$ and $\mathbf{q}_{\rm m2}$ magnetic peaks of domain B are respectively indicated by open and closed blue circles.
}
	\label{separate}
\end{figure*} 
\section{Methods}
\subsection{Experimental Details} 

Single crystals of {\KFAT} were grown using the modified Bridgman method \cite{Lei2011}. \ys{K pieces, Fe powder, Ag powder, and Te powder were mixed with a molar ratio of K:Fe:Ag:Te~=~1:0.8:1.2:2 and loaded into a quartz ampoule. The ampoule was evacuated, sealed, and then further sealed in a larger quartz ampoule. The double-sealed ampoule was placed in a tube furnace with a controlled temperature gradient. The furnace was heated to 750~$^\circ$C and held for 8~h, then heated to 1000~$^\circ$C. During crystal growth, the low-temperature zone was first cooled to 920~$^\circ$C at 20~$^\circ$C/h and then slowly cooled to 420~$^\circ$C at a rate of 3~$^\circ$C/h, while the high-temperature zone was slowly cooled to 500~$^\circ$C at about 3~$^\circ$C/h. The furnace was then turned off and cooled naturally to room temperature.}

Time-of-flight neutron diffraction measurements were carried out using the Elastic Diffuse Scattering Spectrometer (CORELLI) \cite{Ye2018} at the Spallation Neutron Source, Oak Ridge National Laboratory (ORNL). A $\sim3$~g {\KFAT} single crystal was used for the CORELLI experiment, with the $c$-axis oriented vertically. Measurements were carried out via sample rotation around the vertical axis, over 180 degrees in steps of 2 degrees. Polarized neutron scattering measurements were carried out using the Polarized Triple-Axis Spectrometer (HB-1) at the High Flux Isotope Reactor (HFIR), using a $\sim10$~g array of {\KFAT} single crystals co-aligned in the $[HHL]_{\rm T}$ scattering plane. Access to momenta slightly out of this plane was achieved by tilting the goniometer of the sample stage. Synchrotron X-ray diffraction (XRD) measurements were performed at the BL02B1 beamline of SPring-8, using X-rays with a wavelength of 0.2880~{\AA}. The beam size was $200\times200$~$\mu$m$^2$, and the sample was smaller than $50 \times 50 \times 50~\mu\mathrm{m}^3$.

\ys{Fe $L_{2,3}$-edge X-ray absorption spectroscopy (XAS) measurements were performed on {\KFAT} and BaFe$_2$As$_2$ at the BL07U beamline of the Shanghai Synchrotron Radiation Facility (SSRF) using linearly polarized X-rays over an energy range of 698.7--738.7~eV, covering the Fe-$L_{2,3}$ edge. Energy calibration was performed using an Fe$_2$O$_3$ reference prior to sample measurements. The spectra were collected in total electron yield (TEY) mode under ultrahigh vacuum conditions. Single-crystal samples were cleaved in situ in vacuum before measurements. The energy resolution at the Fe $L_{2,3}$ edge was approximately 0.1~eV. A Shirley background subtraction was applied to remove the background contribution from the spectra \cite{HerreraGomez2012}. Fe $K$-edge XAS measurements were performed on {\KFAT} single crystals at 3.4~K at the BL39XU beamline of SPring-8 in fluorescence-yield mode using linearly polarized X-rays over an energy range of 7100--7160~eV. The single crystals were cleaved in situ under vacuum at low temperature to expose fresh surfaces prior to the measurements. The incident X-ray intensity $I_0$ was monitored using an ionization chamber, and the Fe fluorescence signal was collected using a seven-element silicon drift detector (SDD). A Si(220) double-crystal monochromator was used, providing an energy resolution of approximately 0.41~eV at 7.14~keV. The energy was calibrated using an Fe foil reference, and the incident beam size at the sample position was approximately $0.5\times0.5$~mm$^2$. A linear pre-edge background fitted over 7100--7107~eV was subtracted from each repeated scan. The background-subtracted spectra were then normalized by their respective mean intensities over 7140--7160~eV.} The optical reflectivity measurements were performed on a Bruker Vertex 80V spectrometer, with the sample coated in a gold or aluminum film.

\subsection{Data Integration and Refinement}
For the CORELLI data, the Bragg peaks were indexed and peak intensities were integrated for each individual structural and magnetic domain. Peaks overlapping with aluminum powder rings were excluded from the analysis. The refinement was performed using the \texttt{Fullprof Suite} \cite{fullprof}. For the superstructure refinement, the main structural reflections were excluded because they were shared by the two superstructure twin domains. For the magnetic refinement, nearly overlapping reflections were considered to yield unreliable integrated intensities and were therefore excluded. Eventually, only one or two parameters were used to describe the magnetic structure that was refined against 150-200 magnetic reflections, as detailed later in Sec. III F.



\begin{table}[t!]
    \centering
    \caption{\ys{Indexing relations between the $I4/mmm$ parent cell and the
    $I4$ superstructure cells.}}
    \label{tab:cell_conversion}

    \renewcommand{\arraystretch}{1.20}
    \setlength{\tabcolsep}{2.5pt}
    \scriptsize

    \begin{tabular}{|c|c|c|}
        \hline

        \parbox[c][3.2em][c]{0.24\columnwidth}{
            \centering
            \textbf{$I4/mmm$ cell}
        }
        &
        \multicolumn{2}{c|}{
            \parbox[c][3.2em][c]{0.67\columnwidth}{
                \centering
                \textbf{$I4$ superstructure cell}
            }
        }
        \\
        \hline

        \parbox[c][8.2em][c]{0.24\columnwidth}{
            \centering
            $a_{\mathrm T}=b_{\mathrm T}
            \approx 4.37~\text{\AA}$\\[6pt]
            $c_{\mathrm T}\approx 14.95~\text{\AA}$
        }
        &
        \multicolumn{2}{c|}{
            \parbox[c][8.2em][c]{0.67\columnwidth}{
                \centering
                $a_{\mathrm S}=b_{\mathrm S}
                =\sqrt{5}\,a_{\mathrm T}
                =\sqrt{5}\,b_{\mathrm T}$,\\[4pt]
                $c_{\mathrm S}=c_{\mathrm T}$\\[6pt]
                magnetic propagation vectors:\\[4pt]
                $\mathbf{q}_{\mathrm{m1}}
                =(0.29,-0.26,0)_{\mathrm S}$\\[4pt]
                $\mathbf{q}_{\mathrm{m2}}
                =(0.26,0.29,0)_{\mathrm S}$
            }
        }
        \\
        \hline

        \parbox[c][12.5em][c]{0.24\columnwidth}{
            \centering
            arbitrary vector in\\[2pt]
            reciprocal space:\\[6pt]
            $\displaystyle
           \mathbf{Q}_{\mathrm T}
            =
            \begin{pmatrix}
                H\\
                K\\
                L
            \end{pmatrix}_{\mathrm T}$
        }
        &
        \parbox[c][12.5em][c]{0.335\columnwidth}{
            \centering
            \textbf{SS domain A}\\[6pt]
            {\footnotesize
            \begingroup
            \setlength{\arraycolsep}{1.8pt}
            \(
            \begin{pmatrix}
                H\\
                K\\
                L
            \end{pmatrix}_{\mathrm{SA}}
            =
            \begin{pmatrix}
                 2 & -1 & 0\\
                 1 &  2 & 0\\
                 0 &  0 & 1
            \end{pmatrix}
            \begin{pmatrix}
                H\\
                K\\
                L
            \end{pmatrix}_{\mathrm T}
            \)
            \endgroup
            }
        }
        &
        \parbox[c][12.5em][c]{0.335\columnwidth}{
            \centering
            \textbf{SS domain B}\\[6pt]
            {\footnotesize
            \begingroup
            \setlength{\arraycolsep}{1.8pt}
            \(
            \begin{pmatrix}
                H\\
                K\\
                L
            \end{pmatrix}_{\mathrm{SB}}
            =
            \begin{pmatrix}
                 2 & 1 & 0\\
                -1 & 2 & 0\\
                 0 & 0 & 1
            \end{pmatrix}
            \begin{pmatrix}
                H\\
                K\\
                L
            \end{pmatrix}_{\mathrm T}
            \)
            \endgroup
            }
        }
        \\


        \parbox[c][12.5em][c]{0.24\columnwidth}{
            \centering
            direct lattice\\[2pt]
            basis vectors:\\[6pt]
            $(\mathbf{a}_{\mathrm T},
              \mathbf{b}_{\mathrm T},
              \mathbf{c}_{\mathrm T})$
        }
        &
        \parbox[c][12.5em][c]{0.335\columnwidth}{
            \centering
            {\footnotesize
            \begingroup
            \setlength{\arraycolsep}{2pt}
            \(
            \begin{gathered}
                (
                \mathbf{a}_{\mathrm{SA}},
                \mathbf{b}_{\mathrm{SA}},
                \mathbf{c}_{\mathrm{SA}}
                )
                \\[6pt]
                =
                (
                \mathbf{a}_{\mathrm T},
                \mathbf{b}_{\mathrm T},
                \mathbf{c}_{\mathrm T}
                )
                \\[4pt]
                {}\times
                \begin{pmatrix}
                     2 & 1 & 0\\
                    -1 & 2 & 0\\
                     0 & 0 & 1
                \end{pmatrix}
            \end{gathered}
            \)
            \endgroup
            }
        }
        &
        \parbox[c][12.5em][c]{0.335\columnwidth}{
            \centering
            {\footnotesize
            \begingroup
            \setlength{\arraycolsep}{2pt}
            \(
            \begin{gathered}
                (
                \mathbf{a}_{\mathrm{SB}},
                \mathbf{b}_{\mathrm{SB}},
                \mathbf{c}_{\mathrm{SB}}
                )
                \\[6pt]
                =
                (
                \mathbf{a}_{\mathrm T},
                \mathbf{b}_{\mathrm T},
                \mathbf{c}_{\mathrm T}
                )
                \\[4pt]
                {}\times
                \begin{pmatrix}
                    2 & -1 & 0\\
                    1 &  2 & 0\\
                    0 &  0 & 1
                \end{pmatrix}
            \end{gathered}
            \)
            \endgroup
            }
        }
        \\
        \hline

    \end{tabular}
\end{table}

\subsection{\ys{Relation between $I4/mmm$ and $I4$ cells}}
When the Fe-Ag ordering in {\KFAT} is ignored, it has the same $I4/mmm$ main structure as BaFe$_2$As$_2$, with two (Fe,Ag)-Te layers and 4 Fe/Ag atoms (2 per layer) in the tetragonal unit cell. The Fe-Ag ordering leads to a chiral $I4$ superstructure with a $\sqrt{5}\times\sqrt{5}$ expansion in the $ab$-plane, resulting in two (Fe,Ag)Te layers with 8 Fe and 12 Ag atoms (10 Fe/Ag per layer) in the expanded tetragonal cell \cite{Song2019}. Throughout this work, references to momentum indexed by the $I4/mmm$ cell are indicated by the subscript `T', and the $I4$ cell by `S'. \ys{Subscripts `SA' or `SB' is used to specifically reference momenta for superstructure domain A or domain B, respectively. The conversion between momenta indexed by the $I4/mmm$ and $I4$ cells are summarized in Tab.~\ref{tab:cell_conversion}. Magnetic peaks occur at ${\bf G}_{\rm  S}\pm {\bf q}_{\rm m1}$ or ${\bf G}_{\rm  S}\pm {\bf q}_{\rm m2}$, where ${\bf G}_{\rm S}$ is a Bragg peak of the $I4$ cell, $\mathbf{q}_{\rm m1}=(0.29,-0.26,0)_{\rm  S}$, and $\mathbf{q}_{\rm m2}=(0.26,0.29,0)_{\rm  S}$. Superstructure peaks are ${\bf G}_{\rm  S}$ peaks that do not coincide with Bragg peaks of the $I4/mmm$ cell.}


\begin{figure}
	\includegraphics[width=1\columnwidth]{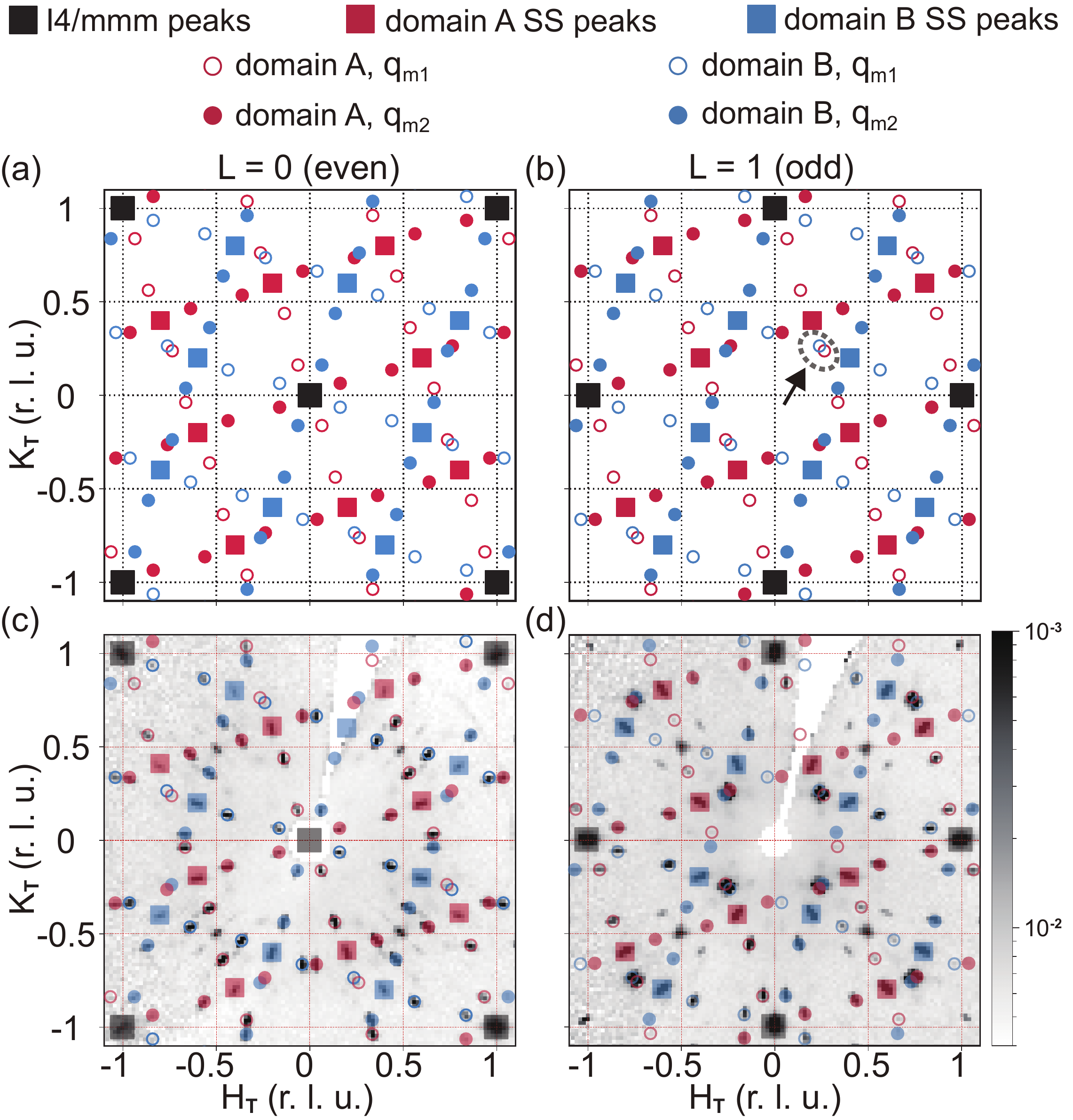} \protect\caption{Schematic scattering maps in reciprocal space when contributions from all domains are considered, for (a) $HK$-planes with even $L$, and (b) $HK$-planes with odd $L$. Single crystal neutron diffraction maps at 7~K from CORELLI, for the (c) $[HK0]_{\rm T}$ and (d) $[HK1]_{\rm T}$ planes, with expected Bragg peaks overlaid on the data.}
	\label{full}
\end{figure}

\begin{figure*}
	\includegraphics[width=2\columnwidth]{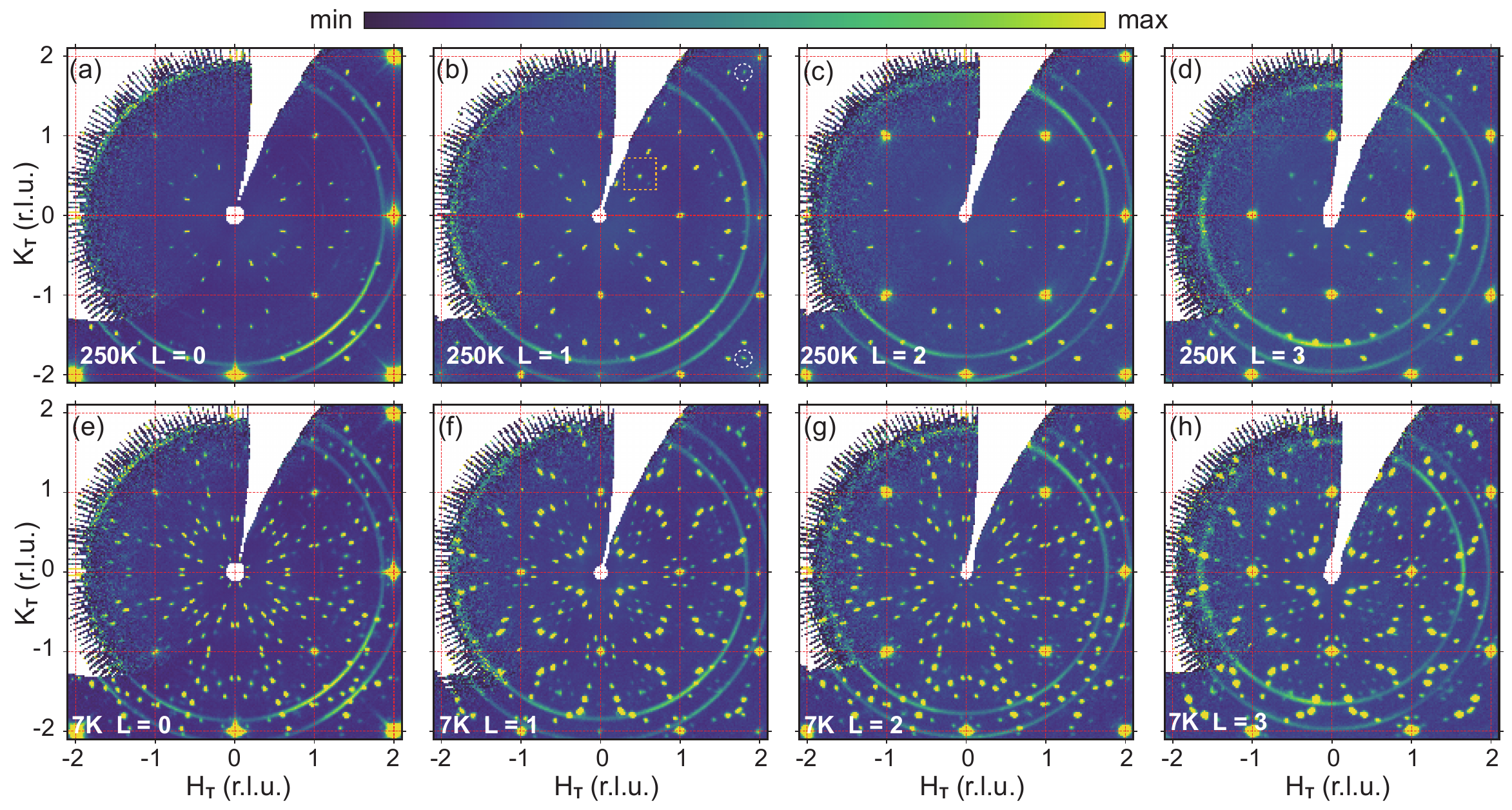} \protect\caption{Neutron diffraction maps of {\KFAT} at 250~K measured using CORELLI, in $HK$-planes with (a) $L$ = 0, (b) $L=1$, (c) $L=2$, and (d) $L=3$. The corresponding $HK$-planes maps at 7~K are shown in (e), (f), (g), and (h), respectively. These maps were obtained by binning data within $L = \pm 0.1$ around the central $L$ value. A large number of magnetic peaks appear at 7~K.}
	
	\label{TOF}
\end{figure*}

\begin{figure}
	\includegraphics[width=1\columnwidth]{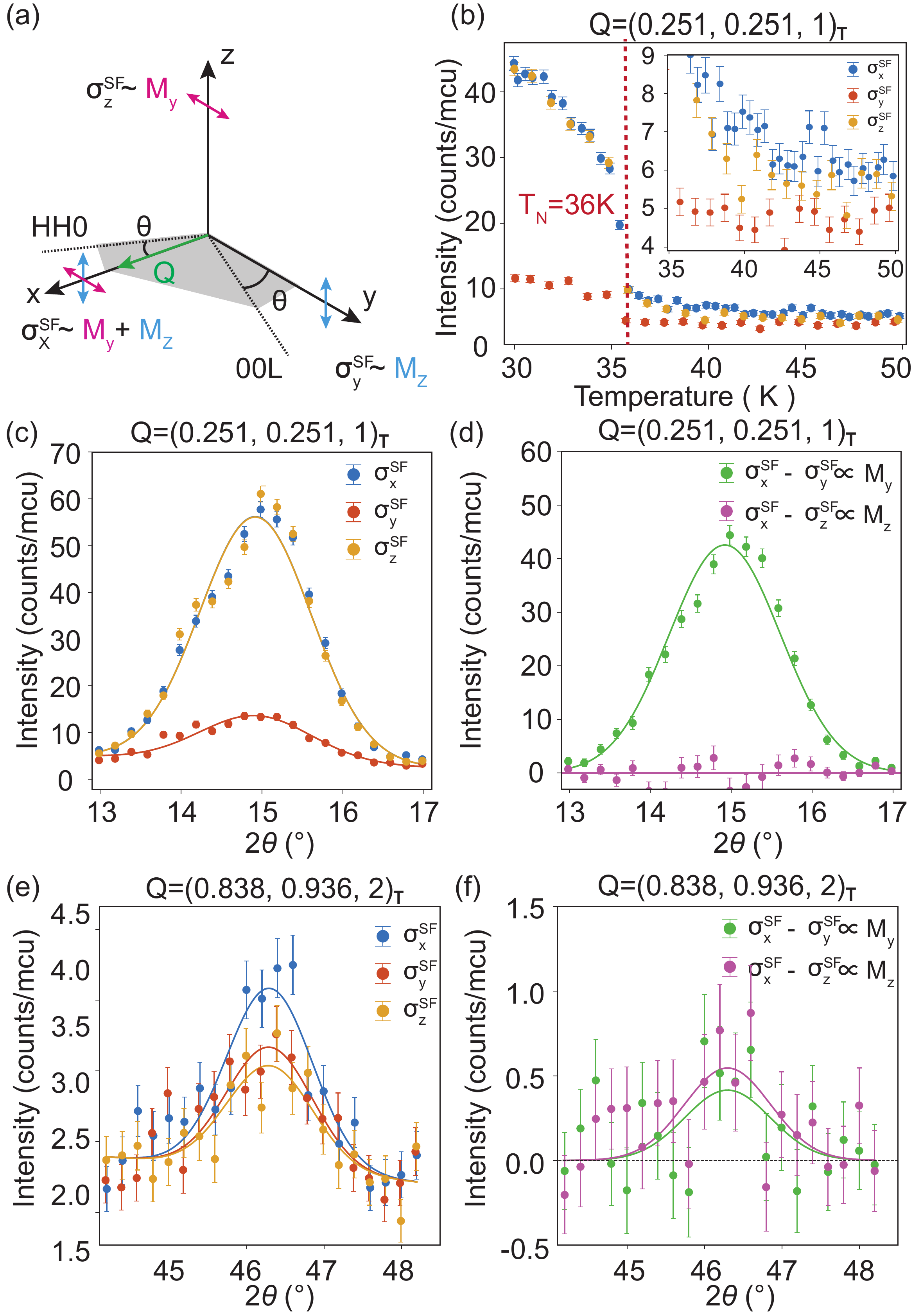} \protect\caption{(a) Schematic of the polarized neutron scattering experiment. (b) Temperature dependence of the magnetic Bragg peak $(0.251,0.251,1)_{\rm T}$, for the three SF channels. $\theta-2\theta$ scans centered around magnetic Bragg peaks at (c) $(0.251,0.251,1)_{\rm T}$, and (e) $(0.838,0.936,2)_{\rm T}$. The corresponding $\sigma_{x}^{\rm SF}-\sigma_{y}^{\rm SF}$ and $\sigma_{x}^{\rm SF}-\sigma_{z}^{\rm SF}$ at these positions are shown in (d) and (f).}
	\label{sf}
\end{figure}

\section{Results}

\subsection{X-ray absorption spectroscopy}

\ys{X-ray absorption spectroscopy (XAS) spectra near the Fe $L_{2,3}$-edge are compared for {\KFAT}, BaFe$_2$As$_2$, and FeTe~\cite{saini2011} in Fig.~\ref{basic}(a). After the subtraction of a Shirley background, each spectrum was normalized to its maximum intensity. The Fe $L_{2,3}$-edge energy and overall spectral profile of {\KFAT} are similar between the shown compounds, as well as other related Fe-based materials such as FeSe \cite{Kim2016} and NaFe$_{0.53}$Cu$_{0.47}$As \cite{Song2021}. 
The spectra are dominated by a main absorption peak near $E=707.9$~eV and show no prominent feature at $E\approx710$~eV, which suggest that Fe ions in {\KFAT} adopt a valence close to Fe$^{2+}$, without evidence for a sizable Fe$^{3+}$ contribution.}

\ys{The Fe $K$-edge XAS spectra of {\KFAT}, BaFe$_2$As$_2$~\cite{Bittar2011CoSubstitution}, and FeTe~\cite{joseph2010}, are compared in Fig.~\ref{basic}(b). The Fe $K$-edge spectrum of {\KFAT} closely resembles that of FeTe, reflecting similar Fe valence and Te-based local coordination. These results further evidence that the Fe valence in {\KFAT} is close to $2+$.}

\ys{Based on the nominal valences of K$^{+}$, Ag$^{+}$, and Te$^{2-}$, stoichiometric {\KFAT} would be expected to have an average Fe valence of Fe$^{2.25+}$. In order to allow for Fe$^{2+}$, we speculate that there may be Te vacancies, or Te may adopt a non-integer valence. Further studies are required to understand this discrepancy.}

\subsection{Optical Conductivity}

Electrical transport and angle-resolved photoemission spectroscopy (ARPES) measurements indicate {\KFAT} to be a semiconductor/insulator, although the transport gap ($\sim0.1$~eV) is much smaller than the gap seen in ARPES ($\sim0.5$~eV below $E_{\rm F}$) \cite{Lei2011,Ang2013}. To further shed light on the bulk electronic gap of {\KFAT}, we measured its room temperature optical reflectivity $R(\omega)$ over a broad frequency range and derived its optical conductivity $\sigma_1(\omega)$ via the Kramers-Kronig transformation, as shown in Fig.~\ref{basic}(c). \ys{ Below $\sim0.12$~eV, $\sigma_1(\omega)$ is vanishingly small except for a series of sharp phonon excitations [black arrows in Fig.~\ref{basic}(c)], consistent with infrared-active vibrations in an insulating lattice.} These observations unequivocally demonstrate that {\KFAT} is a semiconductor/insulator with a sizable bulk gap that exceeds \ys{$\sim0.12$~eV}. 

\ys{Quantitative analysis of the band gap is achieved using the Tauc relation \cite{Tauc1966},
$(\varepsilon_2\omega^2)^n \propto \hbar\omega-E_{\rm g}$,
where $\varepsilon_2$ is the imaginary part of the dielectric function, $\hbar\omega$ is the photon energy, and $E_{\rm g}$ denotes the optical band gap. For a direct gap transition, $n=2$ is expected. As shown in the inset of Fig.~\ref{basic}(c), the corresponding Tauc plot exhibits a well-defined linear region marked by the dashed line, and extrapolation of this linear region gives a direct optical gap of $\approx1.09$~eV.} 

\ys{This value is close to twice of $\sim$0.5~eV, the separation between valence-band maximum and the Fermi level ($E_{\rm F}$) observed via ARPES \cite{Ang2013}. Since optical spectroscopy probes the full band gap between occupied valence-band states and unoccupied conduction-band states, and assuming that $E_{\rm F}$ is located in the middle of the valence-band top and conduction-band bottom, the measured optical gap of $\approx1.09$~eV is in good agreement with ARPES measurements.}

\subsection{Synchrotron X-ray Diffraction}

The $\sqrt{5}\times\sqrt{5}$ structure of {\KFAT} has two domains with identical main Bragg peaks but distinct superstructure peaks, as schematically shown in Fig.~\ref{separate} (domains A and B). Such a twinning of domains is also observed in the 245 iron chalcogenides such as K$_2$Fe$_4$Se$_5$, which also have a $\sqrt{5}\times\sqrt{5}$ superstructure that result from the ordering of Fe vacancies \cite{Bao2011,Wang2011,Ye2011}. Because the crystal structure of {\KFAT} is chiral, domain A or B each consists of two chiral domains that have identical scattering [Figs.~\ref{3Dstructure}(a) and (c)], which correspond to structures in Figs.~\ref{separate}(a)-(d) with Te at the center of $2\times2$ Fe blocks either below or above the shown Fe-Ag layer. Because the two chiral domains have identical signatures in scattering measurements, we do not explicitly distinguish between them in discussions below.

The presence of both A and B domains give rise to octets of superstructure peaks around $H_{\rm T}+K_{\rm T}={\rm odd}$ in the even-$L$ planes, and around $H_{\rm T}+K_{\rm T}={\rm even}$ in odd-$L$ planes, as schematically shown by red and blue square symbols in Figs.~\ref{basic}(d)-(e) and Fig.~\ref{full}. Synchrotron XRD measurements of {\KFAT} in the $[HK0]_{\rm T}$ and $[HK1]_{\rm T}$ planes are shown in Figs.~\ref{basic}(d) and (e), with these superstructure octets clearly observed, consistent with lab XRD diffraction in Ref.~\cite{Song2019}. \ys{The intensities of the superstructure peaks suggest that the A and B domains are equally populated within the $\sim50\times50\times20$~$\mu$m$^3$ sample probed in the synchrotron XRD measurements.}

\subsection{Time-of-flight Elastic Neutron Scattering}

Well above $T_{\rm N}$ at 250~K, the $\sqrt{5}\times\sqrt{5}$ superstructure peaks are also clearly observed in neutron diffraction $[HK]_{\rm T}$ maps with various $L$ [Figs.~\ref{TOF}(a)-(d)]. In both the XRD and neutron diffraction data, there are additional peaks above $T_{\rm N}$ that cannot be indexed by the $\sqrt{5}\times\sqrt{5}$ structure. These peaks are likely associated with intergrown phases, and are further discussed below and in the Appendix.

Below $T_{\rm N}\approx T_{\rm nem}$, the coupled magnetic and nematic orders break fourfold rotation symmetry of the $I4$ cell, and their coupling to the lattice leads to a structural distortion. The lattice is elongated along one Fe-Fe direction and contracted along the other, which generates two nematic domains for both superstructure domains A [gray arrows in Figs.~\ref{separate}(a) and (b)] and B [gray arrows in Figs.~\ref{separate}(c) and (d)]. The structural distortion results in nematic domains that manifest as a small splitting or broadening of structural Bragg peaks, which were examined in previous works \cite{Song2019,Giles-Donovan2025}. In scattering data presented in this work, we ignore effects due to the structural transition at $T_{\rm nem}$, as they are negligible at relatively small $|\mathbf{Q}|$. 

The magnetic order in {\KFAT} is coupled to the nematic order, and is characterized by a single propagation vector, either $\mathbf{q}_{\rm m1}=(0.29,-0.26,0)_{\rm S}$ or $\mathbf{q}_{\rm m2}=(0.26,0.29,0)_{\rm  S}$. In each $I4$ primitive unit cell of {\KFAT}, there are 4 Fe atoms that form a $2\times2$ block, which are modulated by $\mathbf{q}_{\rm m 1}$ or $\mathbf{q}_{\rm m 2}$ from block to block. Within each block, the Fe atoms adopt a stripe-type configuration, as schematically depicted by ‘+’ and '-' symbols in Figs.~\ref{separate}(a)-(d). Importantly, the direction of FM (AFM) alignment is also the direction of lattice contraction (elongation) \cite{Song2019}, demonstrating a coupling between nematicity and stripe magnetic order similar to the FeSCs \cite{Dai2015}. Thus, the magnetic/nematic domains could be categorized into two types: $\mathbf{q}_{\rm m1}$ domains with elongated AFM Fe-Fe bonds along $(110)_{\rm T}$ and contracted FM Fe-Fe bonds along $(1\overline{1}0)_{\rm T}$ [Figs.~\ref{separate}(a) and (c)], and $\mathbf{q}_{\rm m2}$ domains with elongated AFM Fe-Fe bonds along $(1\overline{1}0)_{\rm T}$ and contracted FM Fe-Fe bonds along $(110)_{\rm T}$ [Figs.~\ref{separate}(b) and (d)]. By applying uniaxial strain, it is possible to detwin {\KFAT} such that either $\mathbf{q}_{\rm m1}$ or  $\mathbf{q}_{\rm m2}$ domains become dominant in the sample \cite{Song2019a}. 

In the FeSCs, stripe magnetism domains that are AFM along $(110)_{\rm T}$ and FM along $(1\overline{1}0)_{\rm T}$ order at $(\frac{1}{2},\frac{1}{2})_{\rm T}$, while domains that are AFM along $(1\overline{1}0)_{\rm T}$ and FM along $(110)_{\rm T}$ order at $(\frac{1}{2},\overline{\frac{1}{2}})_{\rm T}$ . 
Due to the intra-block stripe-like configuration in {\KFAT}, the magnetic structure factor that modulates the intensities of the $\mathbf{q}_{\rm m1}$ magnetic peaks is maximized around $(\frac{1}{2},\frac{1}{2})_{\rm T}$, while that for the $\mathbf{q}_{\rm m2}$ peaks is maximized around $(\frac{1}{2},\overline{\frac{1}{2}})_{\rm T}$. 
Given the stripe-type configuration within each Fe block, the short-range properties of $\mathbf{q}_{\rm m1}$ ($\mathbf{q}_{\rm m2}$) domains in KFAT are expected to resemble the $(\frac{1}{2},\frac{1}{2})_{\rm T}$ ($(\frac{1}{2},\overline{\frac{1}{2}})_{\rm T}$) domains in the FeSCs, even though the long-range ordering vectors are rather different.

For each magnetic/nematic domain, the diffraction pattern is relatively simple, with a pair of magnetic peaks appearing around each allowed structural Bragg peak of the $I4$ structure [Figs.~\ref{separate}(e)-(l)]. Because the structure is body-centered, there is a distinction between the even- and odd-$L$ planes. Typical crystals of {\KFAT} are twinned, such that the 4 domains depicted in Figs.~\ref{separate}(a)-(d) are all present, which leads to rather complicated neutron diffraction patterns [Fig.~\ref{full}]. Figs.~\ref{TOF}(e)-(h) shows neutron scattering $(HK)_{\rm T}$ maps at 7~K for several $L$ values, revealing a series of new peaks relative to 250~K. Comparing these neutron scattering maps with Fig.~\ref{full} shows that these new peaks are well accounted for by magnetic peaks of the 4 types of domains. See Figs.~\ref{full}(c)-(d) for direct comparisons. We note that the neutron diffraction maps exhibit an apparent fourfold rotational symmetry that results from \ys{the twinning of equally populated magnetic/nematic domains shown in Figs.~\ref{separate}(a)-(d), despite the scattering pattern of each individual magnetic/nematic domain being twofold symmetric [Figs.~\ref{separate}(e)-(l)]. This is consistent with previous inelastic X-ray scattering measurements, which observed equally populated nematic domains in a much smaller sample \cite{Giles-Donovan2025}.}

\begin{figure*}
	\includegraphics[width=2\columnwidth]{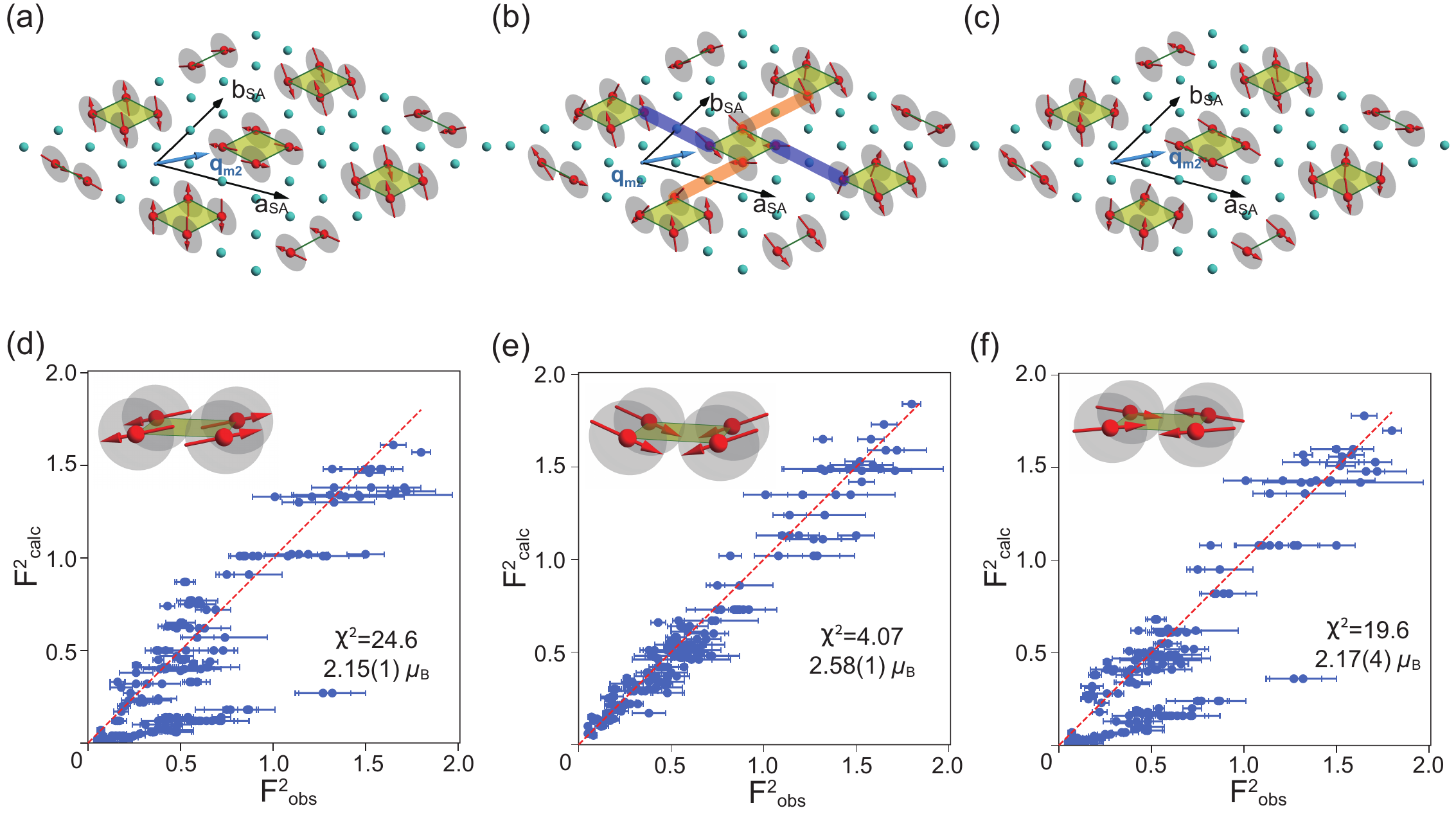} \protect\caption{Candidate magnetic structures considered in this work include (a) collinear spins within each Fe block, (b) collinear spins along the FM direction and non-collinear spins along the AFM direction, and (c) collinear spins along the AFM direction and non-collinear spins along the FM direction. The red and cyan spheres represent Fe and Ag atoms, respectively. The observed and calculated structure factors ($F_{\rm obs}^2$ and $F_{\rm calc}^2$) for these models after refinement are respectively shown in (d), (e), and (f).  
}
	\label{cal_MS}
\end{figure*}
\subsection{Spin anisotropy of the magnetic order}

Magnetic susceptibility of detwinned KFAT suggests the magnetic moments rotate in a plane spanned by the $c$-axis and the elongated AFM Fe-Fe bond direction \cite{Song2019a}. Such a plane was also inferred from magnetic structure refinement, although a slightly rotated plane, such as the plane spanned by the $c$-axis and the superstructure crystal \ys{in-plane} axis that is closer to the AFM Fe-Fe bond direction, also led to a reasonable refinement \cite{Song2019}. To further characterize the spin anisotropy and the plane of spin rotation in {\KFAT}, polarized neutron scattering measurements were carried out, with the results summarized in Fig.~\ref{sf}. 


Figs.~\ref{sf}(a) is a schematic of a typical polarized neutron scattering setup, where the three spin-flip (SF) scattering cross sections $\sigma_{i}^{\rm SF}$ ($i=x,y,z$) are measured \cite{Song2013,Zhang2014}. Following the standard convention, $x\parallel \mathbf{Q}$, $y \perp \mathbf{Q}$ within the scattering plane, and $z$ is perpendicular to the scattering plane. Because SF neutron scattering probes components of the magnetization perpendicular to both $\mathbf{Q}$ and the neutron polarization direction, $\sigma_{x}^{\rm SF}\sim M_y+M_z+B$, $\sigma_y^{\rm SF}\sim M_z+B$, and $\sigma_z^{\rm SF}\sim M_y+B$. $M_i$ ($i=x,y,z$) is the magnetic signal polarized along the $i$-axis, and $B$ is background scattering that is independent of the polarization. In order to eliminate $B$, it is often helpful to examine $\sigma_{x}^{\rm SF}-\sigma_{y}^{\rm SF}\sim M_y$ and $\sigma_{x}^{\rm SF}-\sigma_{z}^{\rm SF}\sim M_z$. We further use $M_{\rm AFM}$, $M_{\rm FM}$, and $M_c$, to respectively denote the magnetic signal polarized along the elongated AFM Fe-Fe direction, the contracted FM Fe-Fe direction, and the $c$-axis.

Figs.~\ref{sf}(c) shows $\sigma_{i}^{\rm SF}$ measured at ${\bf Q}=(0.251,0.251,1)_{\rm T}$, \ys{which probes two almost overlapping $\mathbf{q}_{\rm m1}$ magnetic peaks [\ys{arrow and gray shaded ellipse in} Fig.~\ref{full}(b)]}. Specifically, these are the $\mathbf{Q}=(011)_{\rm \ys{SA}}+\mathbf{q}_{\rm m1}$ peak of superstructure domain A and the $\mathbf{Q}=(101)_{\rm \ys{SB}}-\mathbf{q}_{\rm m1}$ peak of superstructure domain B [Fig.~\ref{full}(b) and Figs.~\ref{separate}(i) and (k)]. At this position,  strong peaks are observed in the $x$ and $z$ channels, with $\sigma_{x}^{\rm SF}\approx\sigma_{z}^{\rm SF}$. While a weak peak is also observed in $\sigma_{y}^{\rm SF}$, it results from a $1/(R+1)$ contribution of $M_y$ \cite{Zhang2014}, where $R\approx7$ is the flipping ratio determined at a structural Bragg peak. The results in Fig.~\ref{sf}(c) evidence a clear peak in $M_y$, and $M_z\approx0$, which can also be seen in Fig.~\ref{sf}(d). Since $z\parallel(1\overline{1}0)_{\rm \ys{T}}$, which corresponds to the contracted FM direction for $\mathbf{q}_{\rm m1}$ domains, $M_z\approx M_{\rm FM}\approx0$ shows that the magnetic order in KFAT does not have a component along the contracted FM direction, and confirms the plane of spin rotation to be spanned by the $c$-axis and the AFM Fe-Fe direction. 

Fig.~\ref{sf}(e) shows $\sigma_{i}^{\rm SF}$ measured at ${\bf Q}=(0.838,0.936,2)_{\rm T}$, which corresponds to the $\mathbf{Q}=(1,3,2)_{\rm \ys{SA}}-\mathbf{q}_{\rm m2}$ magnetic peak of superstructure domain A [Fig.~\ref{full}(a) and Fig.~\ref{separate}(f)]. The three polarization directions are defined for the $[0.838H,0.936H,L]_{\rm T}$ scattering plane, using the same convention described above. Magnetic peaks are seen in all three polarization channels, with $\sigma_{y}^{\rm SF}\approx\sigma_{z}^{\rm SF}$ and peaks that are roughly half of that in $\sigma_{x}^{\rm SF}$. Since this magnetic peak is associated with $\mathbf{q}_{\rm m2}$, $(110)_{\rm T}$ is the contracted FM direction, and $(1\overline{1}0)_{\rm T}$ is the elongated AFM direction [Fig.~\ref{separate}(b)]. It can then be shown that for this $\mathbf{Q}$, $M_y\approx 0.18M_{\rm FM}+0.82M_c\approx0.82M_c$ and $M_z\approx M_{\rm AFM}$. For a circular magnetic helix in the plane spanned by the AFM Fe-Fe direction and the $c$-axis, which applies to {\KFAT}, it is then expected that $M_y/M_z\approx0.82$. Figs.~\ref{sf}(e) and (f) are consistent with this expectation, although the limited count rate gives $M_y/M_z\approx0.76(2)$,  which suggests possible deviations from a circular magnetic helix. 

Fig.~\ref{sf}(b) shows the temperature dependence of $\sigma_{i}^{\rm SF}$ measured at ${\bf Q}=(0.251,0.251,1)_{\rm T}$, revealing clear quasi-elastic magnetic scattering in the paramagnetic state above $T_{\rm N}$. These quasi-elastic scattering retain a clear spin anisotropy with $M_z\approx M_{\rm FM}\approx0$ above $T_{\rm N}$, identical to that of the magnetic order below $T_{\rm N}$. Such a persistence of spin anisotropy at $\mathbf{q}_{\rm m1}$/$\mathbf{q}_{\rm m2}$ is essential for the in-plane anisotropic uniform (${\bf q}=0$) magnetic susceptibility $\chi$ in uniaxial-strained {\KFAT} \cite{Song2019a}, which we discuss below.

Consider a system with only $\mathbf{q}_{\rm m1}$ magnetic order and fluctuations. Because the helical plane is spanned by the in-plane AFM direction and the $c$-axis, the in-plane easy-axis is $(110)_{\rm T}$. Relative to $(1\overline{1}0)_{\rm T}$, $(110)_{\rm T}$ exhibits a larger $\chi$ above $T_{\rm N}$ and a smaller $\chi$ below $T_{\rm N}$, as expected for a typical easy-axis AFM system. Because unstrained KFAT is tetragonal above $T_{\rm nem}$ ($\sim T_{\rm N}$), $\mathbf{q}_{\rm m1}$ and $\mathbf{q}_{\rm m2}$ fluctuations must be equally present, their anisotropic contributions to $\chi$ balance each other, and results in an in-plane isotropic $\chi$. By applying tensile uniaxial strain along $(110)_{\rm T}$, $\mathbf{q}_{\rm m1}$ fluctuations are favored over $\mathbf{q}_{\rm m2}$ fluctuations, such an imbalance then leads to an in-plane anisotropy in $\chi$, with $\chi_{110}>\chi_{1\overline{1}0}$ above $T_{\rm N}$/$T_{\rm nem}$. Below $T_{\rm N}$/$T_{\rm nem}$, tensile strain along $(110)_{\rm T}$ also favors $\mathbf{q}_{\rm m1}$ magnetic order, which then leads to $\chi_{110}<\chi_{1\overline{1}0}$.

In the scenario described above, there are two necessary ingredients for an in-plane anisotropic $\chi$ in {\KFAT} above $T_{\rm N}$: (1) an imbalance between $\mathbf{q}_{\rm m1}$ and $\mathbf{q}_{\rm m2}$ fluctuations induced by strain, and (2) an in-plane spin anisotropy associated with $\mathbf{q}_{\rm m1}$ or $\mathbf{q}_{\rm m2}$ fluctuations. The measurements in Fig.~\ref{sf}(b) show that (2) is fulfilled, with the spin anisotropy characterized by $M_{z}<M_{y}$ maintained up to 50~K. 

Our polarized neutron scattering shows that the spins in {\KFAT} prefer to lie in the plane spanned by the elongated AFM Fe-Fe direction and the $c$-axis, rather than the contracted FM direction, similar to the parent compounds of the FeSCs \cite{Qureshi2012,Song2013}. The persistence of such an anisotropy above $T_{\rm N}$/$T_{\rm nem}$ (does not require strain) accounts for the in-plane anisotropic $\chi$ in strained {\KFAT}, and motivates a direct measurement of the strain-induced imbalance between $\mathbf{q}_{\rm m1}$ and $\mathbf{q}_{\rm m2}$ fluctuations via inelastic neutron scattering, as done for the FeSCs \cite{Lu2014,Song2015,Lu2018,Tam2020}.

\subsection{Magnetic structure refinement}

Polarized neutron scattering measurements evidence that the ordered spins in {\KFAT} rotate in an easy plane spanned by AFM Fe-Fe bond direction and the $c$-axis. This conclusion is independently obtained through refinements of the neutron diffraction data from CORELLI [Fig.~\ref{TOF}], yielding a spin rotation plane very close to the easy plane. We then fixed the \ys{spin rotation} plane to the easy plane spanned by the AFM Fe-Fe bond direction and the $c$-axis, and performed magnetic structure refinements that compare three candidate magnetic structures [Fig.~\ref{cal_MS}]. 

For each $2\times2$ Fe block, the spins are constrained to be (1) all collinear, or (2) collinear along the FM direction, or (3) collinear along the AFM direction. Each of the four Fe spins then rotates from block to block modulated either by ${\bf q}_{\rm m1}$ or ${\bf q}_{\rm m2}$, forming a simple circular helical magnetic structure. The ordered moment is the only refined parameter for model (1), whereas models (2) and (3) each has an additional refined parameter that characterizes the deviation from a collinear alignment along the AFM or the FM direction. 

For these candidate models, the refined structures are shown in Figs.~\ref{cal_MS}(a)-(c), and the corresponding comparisons between observed and calculated structure factors ($F_{\rm obs}^2$ and $F_{\rm calc}^2$) are shown in Figs.~\ref{cal_MS}(d)-(f). As can be seen, while models (1) and (3) yield reasonable agreements between $F_{\rm obs}^2$ and $F_{\rm calc}^2$, the intensities of some peaks are significantly overestimated ($F_{\rm calc}^2>F_{\rm obs}^2$), leading to large values of $\chi^2$ [Figs.~\ref{cal_MS}(d) and (f)]. On the other hand, model (2) is able to correctly account for these overestimated peaks, leading to a much smaller $\chi^2$. The angle between non-collinear AFM spins in model (2) is refined to be $134.4(8)^\circ$ ($180^\circ$ for collinear AFM spins), and the angle between non-collinear FM spins in model (3) is refined to be $14(2)^\circ$ ($0^\circ$ for collinear FM spins). The ordered moment for model (2) is found to be $2.58(1)$~$\mu_{\rm B}$/Fe, with models (1) and (3) yielding similar values. These values are consistent with previous refinements that considered model (1) \cite{Song2019}. 

Our analysis confirms that the magnetic structure of {\KFAT} is helical, with the four Fe atoms inside the $2\times2$ blocks arranged in a stripe-type configuration. The refinements further suggest deviations from a collinear configuration along the AFM direction. We note that model (2) exhibits a net moment within each Fe block (whereas models (1) and (3) do not), which could potentially be tuned by applied fields. Furthermore, the nearest-neighbor inter-block Fe-Fe couplings in model (2) are highly distinct along the orthogonal in-plane directions: it is nearly AFM along the intra-block AFM direction [purple bonds in Fig.~\ref{cal_MS}(b)], and nearly FM along the intra-block FM direction [orange bonds in Fig.~\ref{cal_MS}(b)]. In contrast, the nearest-neighbor inter-block spins are nearly perpendicular for both in-plane directions in models (1) and (3) [Figs.~\ref{cal_MS}(a) and (c)]. \ys{A non-collinear intra-block configuration may result from Dzyaloshinskii–Moriya interactions or the cooperative effect of intra-block and inter-block couplings. Further measurements of the spin excitations in {\KFAT} may shed light on the microscopic interactions that dictate its magnetic structure.}

Interestingly, the uniform magnetic susceptibility of {\KFAT} follows the Curie-Weiss law very nicely with an effective \ys{moment $\mu_{\rm eff}\approx2.90(6)~\mu_{\rm B}$/Fe \cite{Lei2011,Song2019a}. By assuming $g=2$, this value of $\mu_{\rm eff}$ corresponds to $S\approx1$ and a full ordered moment of $\approx2.0~\mu_{\rm B}$/Fe,} which is slightly smaller than the ordered moment obtained from magnetic structure refinement. Alternatively, the Curie-Weiss behavior may arise from the net moment within each Fe block, which based on the ordered moment for model (2) and the angle between non-collinear AFM spins, yields an ordered moment $\mu_{\rm block}\approx4.0(1)$~$\mu_{\rm B}$. Based on the Curie constant measured for {\KFAT} \cite{Song2019a} and assuming that it arises from an effective moment for each block, we obtain  \ys{$\mu_{\rm block}^{\rm eff}=2\mu_{\rm eff}\approx5.8(1)$~$\mu_{\rm B}$} for each block. Assuming the two values are related by $\mu_{\rm block}^{\rm eff}=\sqrt{\mu_{\rm block}(\mu_{\rm block}+g_{\rm block}\mu_{\rm B})}$, a gyromagnetic factor of $g_{\rm block}\approx4.4$ for the block is needed to make the values of $\mu_{\rm block}^{\rm eff}$ (from magnetic susceptibility) and $\mu_{\rm block}$ (from neutron diffraction refinement) consistent. \ys{Further studies (e.g. using a small crystal optimized for magnetic structure refinement) are desired to confirm the magnetic structure proposed here, and to elucidate its connection to} the simple Curie-Weiss magnetic susceptibility above $T_{\rm N}$.

\subsection{Bragg peaks from intergrown phases}

Although the twinning of superstructure and magnetic/nematic domains successfully explain most of the Bragg peaks observed in {\KFAT}, there are also peaks unaccounted for, which we attribute to intergrown phases. 

In the XRD data, additional peaks are systematically observed around $(0.5,0.5)_{\rm T}$ and $(0.25,0.25)_{\rm T}$ positions at 300~K. 
The $(0.5,0.5)_{\rm T}$ peaks are also observed in the neutron scattering measurements, with an additional quartet of peaks observed around $(0.5,0.5)_{\rm T}$ [dashed box in Figs.~\ref{TOF}(b) and similar regions in other panels of Fig.~\ref{TOF}]. These peaks are most visible at small $|\mathbf{Q}|$ and are not observed in the XRD data, which suggest they could be magnetic peaks of an intergrown phase. Furthermore, peaks around $(0.2,0.2)_{\rm T}$ are also systematically observed in neutron diffraction [Fig.~\ref{2Dmap} and dashed circles in Figs.~\ref{TOF}(b)].

The observation of superstructure peaks that cannot be indexed by the $I4$ phase suggests unknown intergrown phases with lattice parameters very close to that of {\KFAT}. These intergrown phases are likely also derived from the BaFe$_2$As$_2$ $I4/mmm$ structure, but with superstructure modulations distinct from that of {\KFAT}. A similar coexistence of intergrown phases is also observed in the alkaline metal iron chalcogenide superconductors \cite{Kazakov2011,Zhao2012,Bao2013,Wang2014,Wang2016_2}. Estimates of lattice parameters for the intergrown phases are discussed in the Appendix.

\section{Conclusion}
We performed systematic synchrotron X-ray and neutron diffraction measurements of the local-moment iron chalcogenide {\KFAT}, and showed that the $\sqrt{5}\times\sqrt{5}$ superstructure and the intertwined magnetic and nematic orders lead to four twin domains, accounting for most of the Bragg peaks observed in neutron diffraction. Additional superstructure or magnetic peaks that cannot be accounted for are ascribed to intergrown phases, which motivates further exploration. Magnetic structure refinement confirms a stripe-type configuration within the $2\times2$ Fe blocks, and allowing for a noncollinear alignment along the AFM direction leads to an improved refinement over the earlier reported collinear model \cite{Song2019}. Polarized neutron scattering measurements indicate an easy plane spanned by the AFM Fe-Fe bond direction and the $c$-axis, similar to the spin anisotropy in BaFe$_2$As$_2$ and NaFeAs. The persistence of such spin anisotropy above $T_{\rm N}$/$T_{\rm nem}$ is essential for the strain-induced in-plane anisotropy in the uniform magnetic susceptibility. These results demonstrate {\KFAT} to be an insulating analogue of the parent compounds of FeSCs, with its essential physics arising from localized electronic degrees of freedom within $2\times2$ Fe blocks.


\section{Acknowledgments} 
The work acknowledges support from the National Key R\&D Program of China (No.~2022YFA1402200, No.~2023YFA1406500), the National Natural Science Foundation of China (No.~12350710785, No. 12274363), and the Fundamental Research Funds for the Central Universities (Grant No. 226-2024-00068). Work at the University of California, Berkeley and Lawrence Berkeley National Laboratory was funded by the U.S. DOE, Office of Science, Office of Basic Energy Sciences, Materials Sciences and Engineering Division under Contract No. DE-AC02-05CH11231 (Quantum Materials Program KC2202). A portion of this research used resources at The Spallation Neutron Source and the High Flux Isotope Reactor, DOE Office of Science User Facilities operated by the Oak Ridge National Laboratory. The beam times were allocated to CORELLI on proposal number IPTS-22353.1, and to PTAX (HB-1) on proposal number IPTS-25042.1. \ys{The Fe $K$-edge XAS measurements were performed at BL39XU of SPring-8 under proposal No.~2025A1588. We thank Kotaro Higashi and the BL39XU beamline staff for their support during the experiment.}

\section{Appendix}
\subsection{Chiral domains of {\KFAT}}

\begin{figure}
	\includegraphics[width=1\columnwidth]{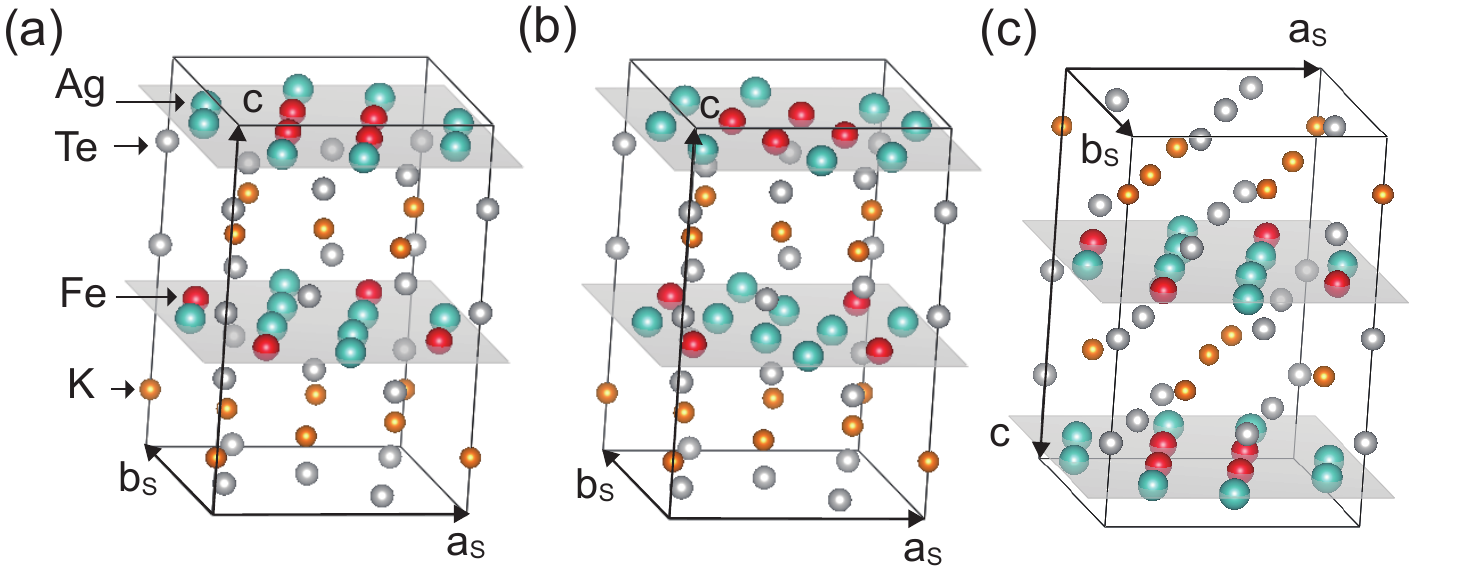} \protect\caption{(a), (b) Two possible chiral variants of {\KFAT}, which are mirror images of one another. (c) Same as (b), but rotated such that the Fe-Ag plane resembles (a). Note that the Te atoms at the center of $2\times2$ Fe blocks are above the Fe-Ag planes in (a), and are below the Fe-Ag planes in (c).}
	\label{3Dstructure}
\end{figure}

Figs.~\ref{3Dstructure}(a) and (b) show the two chiral crystal structures of {\KFAT} in three dimensions, which are mirror images of each other. Each of the four possible domains shown in Figs.~\ref{separate}(a)-(d) consists of two chiral domains [Figs.~\ref{3Dstructure}(a) and (c)], which have Bragg peaks at identical positions [Figs.~\ref{separate}(e)-(l)]. These overlapping chiral domains are distinguished by whether Te atoms at the center of the $2\times2$ Fe blocks are above or below the Fe-Ag layers. 


\subsection{Additional Bragg peaks due to intergrown phases}

\begin{figure}[t!]
	\includegraphics[width=1\columnwidth]{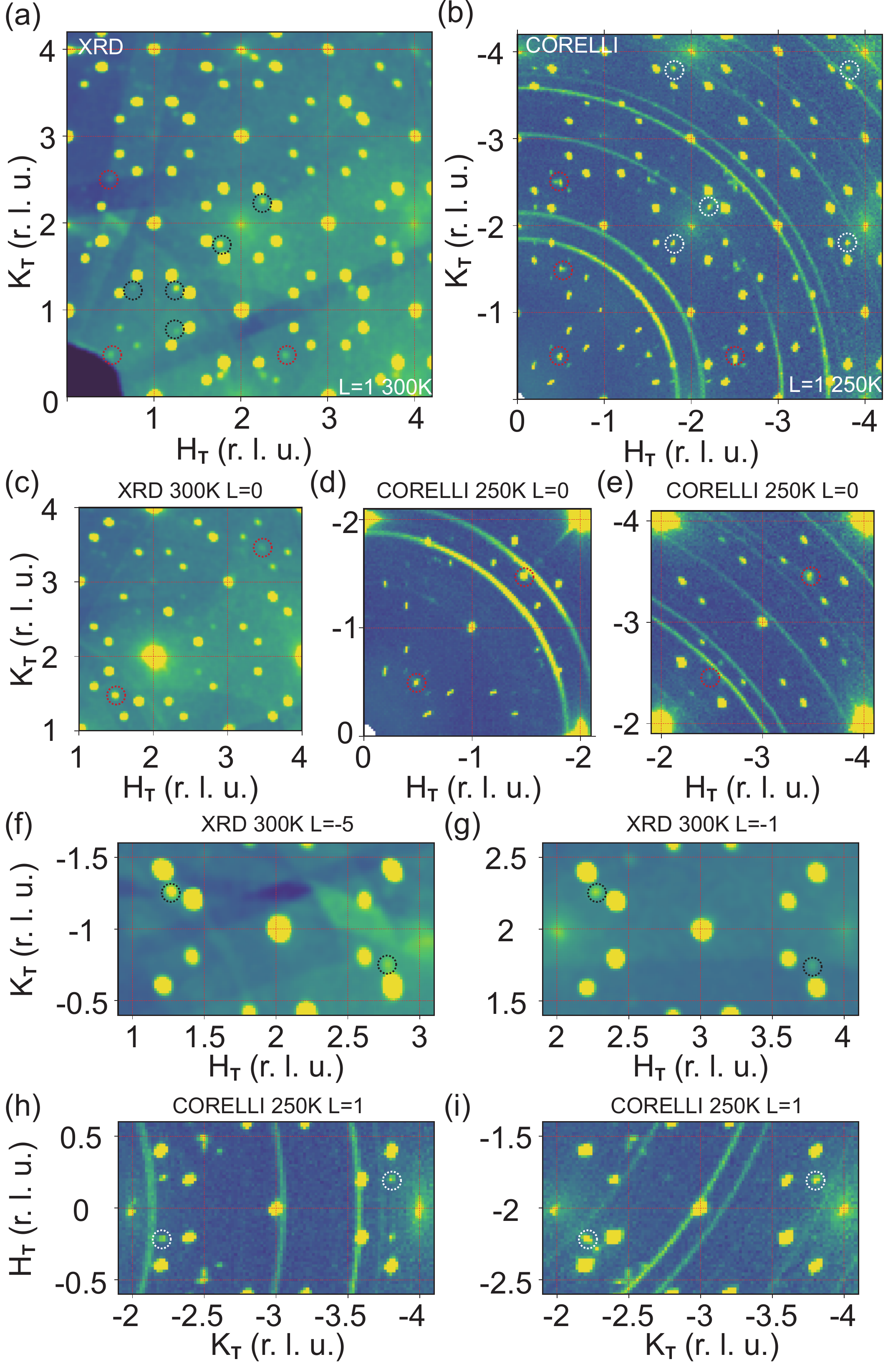} \protect\caption{$[HK1]_{\rm T}$ maps of {\KFAT} measured by (a) XRD at 300~K and (b) neutron diffraction at 250~K. Bragg peaks that cannot be consistently indexed by the superstructure of {\KFAT} are indicated by red dashed circles for peaks around $(0.5,0.5)_{\mathrm{T}}$, black dashed circles for peaks around $(0.25,0.25)_{\mathrm{T}}$, and white dashed circles for peaks around $(0.2,0.2)_{\mathrm{T}}$. These peaks systematically appear across multiple Brillouin zones. Representative $(0.5,0.5)_{\rm T}$ peaks are shown in (c)-(e), $(0.25,0.25)_{\rm T}$ peaks in (f) and (g), and $(0.2,0.2)_{\rm T}$ peaks in (h) and (i).
  }
	\label{2Dmap}
\end{figure}

In Figs.~\ref{2Dmap}(a) and (b), weak Bragg peaks that cannot be indexed by the crystal and magnetic structures of {\KFAT} are indicated by  dashed circles, which we ascribe to intergrown phases. These additional Bragg peaks are weak compared to the $\sqrt{5}\times\sqrt{5}$ superstructure peaks, which suggests that the intergrown phases are present only in trace amounts. 

As intergrown phases typically have lattice parameters close to that of the main phase, their main Bragg peaks often overlap, making it challenging to extract the lattice parameters of intergrown phases. Alternatively, by measuring a pair of superstructure or magnetic peaks of an intergrown phase at ${\bf Q}_1={\bf G}_1+{\bf q}$ and ${\bf Q}_2={\bf G}_2-{\bf q}$, it is possible to extract the lattice spacing of ${\bf G}_1+{\bf G}_2={\bf Q}_1+{\bf Q}_2$ for the intergrown phase, from which its lattice parameters can be estimated.    

Several pairs of ${\bf Q}_1$ and ${\bf Q}_2$ are identified for the $(0.5,0.5)_{\rm T}$ [Figs.~\ref{2Dmap}(c)-(e)], $(0.25,0.25)_{\rm T}$ [Figs.~\ref{2Dmap}(f) and (g)], and $(0.2,0.2)_{\rm T}$ peaks [Figs.~\ref{2Dmap}(h) and (i)]. From these pairs of peaks, we estimate that the $(0.5,0.5)_{\rm T}$, $(0.25,0.25)_{\rm T}$, and $(0.2,0.2)_{\rm T}$ peaks are respectively associated with intergrown phases with in-plane parameters $1.0(6)$\% larger, $0.5(3)$\% smaller, and $0.4(2)$\%~smaller, than $a_{\rm T}\approx4.37$~{\AA} of {\KFAT}.

\begin{figure}[H]
    \centering
    \vspace*{2.0em}
    \includegraphics[width=1\columnwidth]{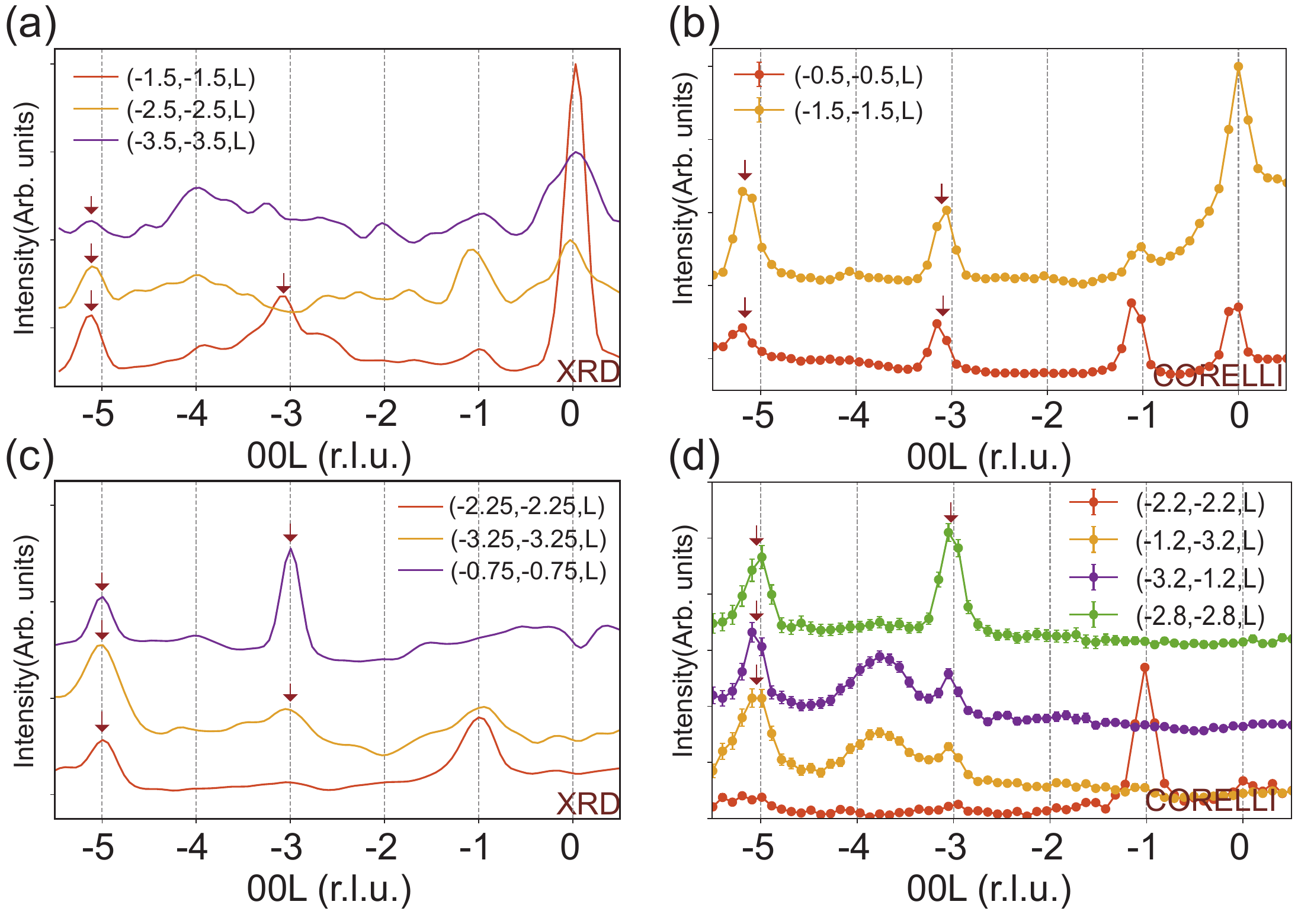}
    \caption{$L$-cuts for (a)--(b) $(0.5,0.5)_{\rm T}$, (c) $(0.25,0.25)_{\rm T}$, and (d) $(0.2,0.2)_{\rm T}$ peaks due to intergrown phases. (a) and (c) are from XRD measurements at 300~K. (b) and (d) are from neutron scattering measurements at 250~K.}
    \label{1Dcuts_L}
\end{figure}

The $c$-axis lattice parameters associated with these weak Bragg peaks are estimated by examining their $L$-dependence. When indexed using the lattice parameters of {\KFAT}, peaks due to intergrown phases will gradually deviate from integer values at large $L$, due to small differences in the $c$-axis lattice parameter. Fig.~\ref{1Dcuts_L} summarizes cuts along $L$ for $(0.5,0.5)_{\rm T}$, $(0.25,0.25)_{\rm T}$, and $(0.2,0.2)_{\rm T}$ peaks. The $(0.5,0.5)_{\rm T}$ [Figs.~\ref{1Dcuts_L}(a) and (b)] and $(0.2,0.2)_{\rm T}$ [Figs.~\ref{1Dcuts_L}(d)] peaks gradually move to larger $L$ positions , which suggests that their associated intergrown phases have slightly smaller lattice parameters than {\KFAT}. No systematic deviations are seen for $(0.25,0.25)_{\rm T}$ peaks [Fig.~\ref{1Dcuts_L}(c)], which suggests that its associated intergrown phase has a $c$-axis lattice parameter very similar to that of {\KFAT}. The lattice parameters of intergrown phases associated with the $(0.5,0.5)_{\rm T}$, $(0.25,0.25)_{\rm T}$, and $(0.2,0.2)_{\rm T}$ peaks are summarized in Tab.~\ref{tab:intergrowth_lattice}. 

\begin{table}[H]
    \centering
    \caption{Estimated lattice parameters of intergrown phases relative to $a_{\rm T}$ and $c_{\rm T}$ of {\KFAT}.}
    \label{tab:intergrowth_lattice}
    \scriptsize
    \renewcommand{\arraystretch}{1.05}

    \begin{tabular*}{\columnwidth}{@{\extracolsep{\fill}}ccc@{}}
    \hline
    \shortstack{Superstructure peaks\\due to intergrown phases} 
    & $a$ 
    & $c$  \\
    \hline
    neutron \& XRD $(0.5,0.5)_{\rm T}$ 
    & $1.010(6)a_{\rm T}$ 
    & $0.974(8)c_{\rm T}$ \\
    XRD $(0.25,0.25)_{\rm T}$  
    & $0.995(3)a_{\rm T}$  
    & $1.00(1)c_{\rm T}$\\
    neutron $(0.2,0.2)_{\rm T}$    
    & $0.996(2)a_{\rm T}$  
    & $0.992(7)c_{\rm T}$  \\
    \hline
    \end{tabular*}
\end{table}

\bibliography{KFAT}

@article{Bittar2011CoSubstitution,
  author  = {Bittar, E. M. and Adriano, C. and Garitezi, T. M. and
             Rosa, P. F. S. and Mendon{\c{c}}a-Ferreira, L. and
             Garcia, F. and Azevedo, G. de M. and Pagliuso, P. G. and
             Granado, E.},
  title   = {{Co}-Substitution Effects on the {Fe} Valence in the
             {BaFe$_2$As$_2$} Superconducting Compound:
             A Study of Hard {X}-Ray Absorption Spectroscopy},
  journal = {Phys. Rev. Lett.},
  volume  = {107},
  number   = {26},
  pages    = {267402},
  year    = {2011},
  month   = dec,
  doi     = {10.1103/PhysRevLett.107.267402},
  url     = {https://doi.org/10.1103/PhysRevLett.107.267402}
}

@techreport{HerreraGomez2012,
  author      = {Herrera-Gomez, Alberto},
  title       = {The Peak-Shirley Background},
  institution = {CINVESTAV, Unidad Quer{\'e}taro},
  type        = {Internal Report},
  pages        = {1--14},
  year        = {2012},
  note        = {Created August 2011; last updated February 2012},
  url         = {https://cia.qro.cinvestav.mx/~aherrera/reportesInternos/peakShirley.pdf}
}

@article{Tauc1966,
  author  = {Tauc, J. and Grigorovici, R. and Vancu, A.},
  title   = {Optical Properties and Electronic Structure of Amorphous Germanium},
  journal = {Physica Status Solidi B},
  volume  = {15},
  pages    = {627--637},
  year    = {1966},
  doi     = {10.1002/pssb.19660150224}
}

@article{joseph2010,
  title = {A study of the electronic structure of {FeSe}$_{1-x}${Te}$_x$ chalcogenides by {Fe} and {Se} {K}-edge x-ray absorption near edge structure measurements},
  author = {Joseph, B. and Iadecola, A. and Simonelli, L. and Mizuguchi, Y. and Takano, Y. and Mizokawa, T. and Saini, N. L.},
  journal = {Journal of Physics: Condensed Matter},
  volume = {22},
  number  = {48},
  pages  = {485702},
  year = {2010},
  doi = {10.1088/0953-8984/22/48/485702}
}

@article{saini2011,
  title = {Electronic structure of {FeSe}$_{1-x}${Te}$_x$ studied by {Fe} {$L_{2,3}$}-edge x-ray absorption spectroscopy},
  author = {Saini, N. L. and Wakisaka, Y. and Joseph, B. and Iadecola, A. and Dalela, S. and Srivastava, P. and Magnano, E. and Malvestuto, M. and Mizuguchi, Y. and Takano, Y. and Mizokawa, T. and Garg, K. B.},
  journal = {Phys. Rev. B},
  volume = {83},
  number  = {5},
  pages  = {052502},
  year = {2011},
  publisher = {American Physical Society},
  doi = {10.1103/PhysRevB.83.052502}
}

@article{Ye2018,
  title = {Implementation of cross correlation for energy discrimination on the time-of-flight spectrometer {CORELLI}},
  volume = {51},
  ISSN = {1600-5767},
  url = {http://dx.doi.org/10.1107/S160057671800403X},
  doi = {10.1107/S160057671800403X},
  number  = {2},
  journal = {Journal of Applied Crystallography},
  publisher = {International Union of Crystallography (IUCr)},
  author = {Ye,  Feng and Liu,  Yaohua and Whitfield,  Ross and Osborn,  Ray and Rosenkranz,  Stephan},
  year = {2018},
  month = mar,
  pages  = {315--322},
}

@article{Wang2026,
  title = {Quasi-one-dimensional spin excitations in the iron pnictide {NaFe$_{0.53}$Cu$_{0.47}$As}},
	volume = {136},
	ISSN = {1079-7114},
  url = {https://doi.org/10.1103/5vdj-9wlt},
	DOI = {10.1103/5vdj-9wlt},
	number  = {6},
	journal = {Phys. Rev. Lett.},
	publisher = {American Physical Society (APS)},
	author = {Wang, Yifan and Tam, David W. and Wang, Weiyi and Ewings, R. A. and Stewart, J. Ross and Matsuda, Masaaki and Cao, Chongde and Liu, Changle and Yu, Rong and Dai, Pengcheng and Song, Yu},
	year = {2026},
	month = feb,
  pages  = {066503},
}

@article{fullprof,
  title = {Recent advances in magnetic structure determination by neutron powder diffraction},
  volume = {192},
  ISSN = {0921-4526},
  url = {http://dx.doi.org/10.1016/0921-4526(93)90108-I},
  doi = {10.1016/0921-4526(93)90108-I},
  number  = {1--2},
  journal = {Physica B: Condensed Matter},
  publisher = {Elsevier BV},
  author = {Rodr{\'i}guez-Carvajal, Juan},
  year = {1993},
  month = oct,
  pages  = {55--69},
}

@article{Chu2012,
  title = {Divergent Nematic Susceptibility in an Iron Arsenide Superconductor},
  volume = {337},
  ISSN = {1095-9203},
  url = {http://dx.doi.org/10.1126/science.1221713},
  DOI = {10.1126/science.1221713},
  number  = {6095},
  journal = {Science},
  publisher = {American Association for the Advancement of Science (AAAS)},
  author = {Chu,  Jiun-Haw and Kuo,  Hsueh-Hui and Analytis,  James G. and Fisher,  Ian R.},
  year = {2012},
  month = aug,
  pages  = {710--712},
}

@article{Lu2018,
  title = {Spin waves in detwinned {BaFe$_2$As$_2$}},
  author = {Lu, Xingye and Scherer, Daniel D. and Tam, David W. and Zhang, Wenliang and Zhang, Rui and Luo, Huiqian and Harriger, Leland W. and Walker, H. C. and Adroja, D. T. and Andersen, Brian M. and Dai, Pengcheng},
  journal = {Phys. Rev. Lett.},
  volume = {121},
  issue = {6},
  pages  = {067002},
  numpages = {6},
  year = {2018},
  month = {Aug},
  publisher = {American Physical Society},
  doi = {10.1103/PhysRevLett.121.067002},
  url = {https://link.aps.org/doi/10.1103/PhysRevLett.121.067002}
}

@article{Song2015,
  title = {Energy dependence of the spin excitation anisotropy in uniaxial-strained {BaFe$_{1.9}$Ni$_{0.1}$As$_2$}},
  author = {Song, Yu and Lu, Xingye and Abernathy, D. L. and Tam, David W. and Niedziela, J. L. and Tian, Wei and Luo, Huiqian and Si, Qimiao and Dai, Pengcheng},
  journal = {Phys. Rev. B},
  volume = {92},
  issue = {18},
  pages  = {180504},
  numpages = {6},
  year = {2015},
  month = {Nov},
  publisher = {American Physical Society},
  doi = {10.1103/PhysRevB.92.180504},
  url = {https://link.aps.org/doi/10.1103/PhysRevB.92.180504}
}

@article{Lu2014,
	author = {Xingye Lu and J. T. Park and Rui Zhang and Huiqian Luo and Andriy H. Nevidomskyy and Qimiao Si and Pengcheng Dai},
  title = {Nematic spin correlations in the tetragonal state of uniaxial-strained {BaFe$_{2-x}$Ni$_x$As$_2$}},
	journal = {Science},
	volume = {345},
	number  = {6197},
  pages  = {657--660},
	year = {2014},
	doi = {10.1126/science.1251853},
	URL = {https://www.science.org/doi/abs/10.1126/science.1251853},
}

@article{Zhang2014,
  title = {Anisotropic neutron spin resonance in underdoped superconducting {NaFe$_{1-x}$Co$_x$As}},
  author = {Zhang, Chenglin and Song, Yu and Regnault, L.-P. and Su, Yixi and Enderle, M. and Kulda, J. and Tan, Guotai and Sims, Zachary C. and Egami, Takeshi and Si, Qimiao and Dai, Pengcheng},
  journal = {Phys. Rev. B},
  volume = {90},
  issue = {14},
  pages  = {140502},
  numpages = {5},
  year = {2014},
  month = {Oct},
  publisher = {American Physical Society},
  doi = {10.1103/PhysRevB.90.140502},
  url = {https://link.aps.org/doi/10.1103/PhysRevB.90.140502}
}

@article{Song2013,
  title = {In-plane spin excitation anisotropy in the paramagnetic state of {NaFeAs}},
	volume = {88},
	ISSN = {1550-235X},
  url = {https://doi.org/10.1103/PhysRevB.88.134512},
  doi = {10.1103/PhysRevB.88.134512},
	number  = {13},
	journal = {Phys. Rev. B},
	publisher = {American Physical Society (APS)},
	author = {Song, Yu and Regnault, Louis-Pierre and Zhang, Chenglin and Tan, Guotai and Carr, Scott V. and Chi, Songxue and Christianson, A. D. and Xiang, Tao and Dai, Pengcheng},
	year = {2013},
	month = oct,
  pages  = {134512},
}

@article{Kim2016,
  title = {Soft {X}-ray absorption spectroscopy study of multiferroic {Bi}-substituted {Ba$_{1-x}$Bi$_x$Ti$_{0.9}$Fe$_{0.1}$O$_3$}},
	volume = {69},
	ISSN = {1976-8524},
	url = {http://dx.doi.org/10.3938/jkps.69.361},
	DOI = {10.3938/jkps.69.361},
	number  = {3},
	journal = {Journal of the Korean Physical Society},
	publisher = {Korean Physical Society},
	author = {Kim, Hyun Woo and Kim, D. H. and Lee, Eunsook and Seong, Seungho and Kang, J.-S. and Kim, Deok Hyeon and Lee, B. W. and Ko, Y. and Kim, J.-Y.},
	year = {2016},
	month = aug,
	pages  = {361--364}
}

@article{Song2021,
  title = {Spin dynamics in {NaFeAs} and {NaFe$_{0.53}$Cu$_{0.47}$As} probed by resonant inelastic x-ray scattering},
	volume = {103},
	ISSN = {2469-9969},
  url = {https://doi.org/10.1103/PhysRevB.103.075112},
  doi = {10.1103/PhysRevB.103.075112},
	number  = {7},
	journal = {Phys. Rev. B},
	publisher = {American Physical Society (APS)},
	author = {Song, Yu and Wang, Weiyi and Paris, Eugenio and Lu, Xingye and Pelliciari, Jonathan and Tseng, Yi and Huang, Yaobo and McNally, Daniel and Dantz, Marcus and Cao, Chongde and Yu, Rong and Birgeneau, Robert J. and Schmitt, Thorsten and Dai, Pengcheng},
	year = {2021},
	month = feb,
  pages  = {075112},
}

@article{Qureshi2012,
  title = {Local magnetic anisotropy in {BaFe$_2$As$_2$}: A polarized inelastic neutron scattering study},
	volume = {86},
	ISSN = {1550-235X},
  url = {https://doi.org/10.1103/PhysRevB.86.060410},
  doi = {10.1103/PhysRevB.86.060410},
	number  = {6},
	journal = {Phys. Rev. B},
	publisher = {American Physical Society (APS)},
	author = {Qureshi, N. and Steffens, P. and Wurmehl, S. and Aswartham, S. and B\"{u}chner, B. and Braden, M.},
	year = {2012},
	month = aug,
  pages  = {060410(R)},
}

@article{Free2010,
  title = {Low-temperature nuclear and magnetic structures of {La$_2$O$_2$Fe$_2$OSe$_2$} from x-ray and neutron diffraction measurements},
  author = {Free, David G. and Evans, John S. O.},
  journal = {Phys. Rev. B},
  volume = {81},
  issue = {21},
  pages  = {214433},
  numpages = {7},
  year = {2010},
  month = {Jun},
  publisher = {American Physical Society},
  doi = {10.1103/PhysRevB.81.214433},
  url = {https://link.aps.org/doi/10.1103/PhysRevB.81.214433}
}

@article{Freelon2021,
  title = {Nematic fluctuations in iron-oxychalcogenide {Mott} insulators},
  volume = {6},
  ISSN = {2397-4648},
  url = {http://dx.doi.org/10.1038/s41535-020-00302-5},
  DOI = {10.1038/s41535-020-00302-5},
  number  = {1},
  journal = {npj Quantum Materials},
  publisher = {Springer Science and Business Media LLC},
  author = {Freelon,  B. and Sarkar,  R. and Kamusella,  S. and Br\"{u}ckner,  F. and Grinenko,  V. and Acharya,  Swagata and Laad,  Mukul and Craco,  Luis and Yamani,  Zahra and Flacau,  Roxana and Swainson,  Ian and Frandsen,  Benjamin and Birgeneau,  Robert and Liu,  Yuhao and Karki,  Bhupendra and Alfailakawi,  Alaa and Neuefeind,  Joerg C. and Everett,  Michelle and Wang,  Hangdong and Xu,  Binjie and Fang,  Minghu and Klauss,  H.-H.},
  year = {2021},
  month = jan,
  pages  = {4},
}

@article{Du2012,
  title = {Stripelike magnetism in a mixed-valence insulating state of the {Fe}-based ladder compound {CsFe$_2$Se$_3$}},
  author = {Du, Fei and Ohgushi, Kenya and Nambu, Yusuke and Kawakami, Takateru and Avdeev, Maxim and Hirata, Yasuyuki and Watanabe, Yoshitaka and Sato, Taku J and Ueda, Yutaka},
  year = {2012},
  month = jun,
  journal = {Phys. Rev. B},
  volume = {85},
  number  = {21},
  pages  = {214436},
  publisher = {American Physical Society},
  doi = {10.1103/PhysRevB.85.214436},
  urldate = {2025-06-25}
}

@article{Takahashi2015,
  title = {Pressure-induced superconductivity in the iron-based ladder material {BaFe$_2$S$_3$}},
  volume = {14},
  ISSN = {1476-4660},
  url = {http://dx.doi.org/10.1038/nmat4351},
  DOI = {10.1038/nmat4351},
  number  = {10},
  journal = {Nature Materials},
  publisher = {Springer Science and Business Media LLC},
  author = {Takahashi,  Hiroki and Sugimoto,  Akira and Nambu,  Yusuke and Yamauchi,  Touru and Hirata,  Yasuyuki and Kawakami,  Takateru and Avdeev,  Maxim and Matsubayashi,  Kazuyuki and Du,  Fei and Kawashima,  Chizuru and Soeda,  Hideto and Nakano,  Satoshi and Uwatoko,  Yoshiya and Ueda,  Yutaka and Sato,  Taku J. and Ohgushi,  Kenya},
  year = {2015},
  month = jul,
  pages  = {1008--1012},
}

@article{Wang2015,
  title = {Mott localization in a pure stripe antiferromagnet {Rb$_{1-\delta}$Fe$_{1.5-\sigma}$S$_2$}},
	volume = {92},
	ISSN = {1550-235X},
  url = {https://doi.org/10.1103/PhysRevB.92.121101},
  doi = {10.1103/PhysRevB.92.121101},
	number  = {12},
	journal = {Phys. Rev. B},
	publisher = {American Physical Society (APS)},
	author = {Wang, Meng and Yi, Ming and Cao, Huibo and de la Cruz, C. and Mo, S. K. and Huang, Q. Z. and Bourret-Courchesne, E. and Dai, Pengcheng and Lee, D. H. and Shen, Z. X. and Birgeneau, R. J.},
	year = {2015},
	month = sep,
  pages  = {121101(R)},
}

@article{Zhao2012,
  title = {Neutron-diffraction measurements of an antiferromagnetic semiconducting phase in the vicinity of the high-temperature superconducting state of {K$_x$Fe$_{2-y}$Se$_2$}},
  author = {Zhao, Jun and Cao, Huibo and Bourret-Courchesne, E. and Lee, D.-H. and Birgeneau, R. J.},
  journal = {Phys. Rev. Lett.},
  volume = {109},
  issue = {26},
  pages  = {267003},
  numpages = {5},
  year = {2012},
  month = {Dec},
  publisher = {American Physical Society},
  doi = {10.1103/PhysRevLett.109.267003},
  url = {https://link.aps.org/doi/10.1103/PhysRevLett.109.267003}
}

@article{Song2016,
	title = {A {Mott} insulator continuously connected to iron pnictide superconductors},
	author = {Song, Yu and Yamani, Zahra and Cao, Chongde and Li, Yu and Zhang, Chenglin and Chen, Justin S. and Huang, Qingzhen and Wu, Hui and Tao, Jing and Zhu, Yimei and Tian, Wei and Chi, Songxue and Cao, Huibo and Huang, Yao-Bo and Dantz, Marcus and Schmitt, Thorsten and Yu, Rong and Nevidomskyy, Andriy H. and Morosan, Emilia and Si, Qimiao and Dai, Pengcheng},
	year = {2016},
	month = dec,
	journal = {Nature Communications},
	volume = {7},
	number  = {1},
	pages  = {13879},
	publisher = {Nature Publishing Group},
	issn = {2041-1723},
	doi = {10.1038/ncomms13879},
	urldate = {2025-06-25}
}

@article{Singh2008,
  title = {Density functional study of {LaFeAsO$_{1-x}$F$_x$}: A low carrier density superconductor near itinerant magnetism},
  author = {Singh, D. J. and Du, M.-H.},
  journal = {Phys. Rev. Lett.},
  volume = {100},
  issue = {23},
  pages  = {237003},
  numpages = {4},
  year = {2008},
  month = {Jun},
  publisher = {American Physical Society},
  doi = {10.1103/PhysRevLett.100.237003},
  url = {https://link.aps.org/doi/10.1103/PhysRevLett.100.237003}
}

@article{Si2008,
  title = {Strong Correlations and Magnetic Frustration in the High ${T}_{c}$ Iron Pnictides},
  author = {Si, Qimiao and Abrahams, Elihu},
  journal = {Phys. Rev. Lett.},
  volume = {101},
  issue = {7},
  pages  = {076401},
  numpages = {4},
  year = {2008},
  month = {Aug},
  publisher = {American Physical Society},
  doi = {10.1103/PhysRevLett.101.076401},
  url = {https://link.aps.org/doi/10.1103/PhysRevLett.101.076401}
}

@article{Yin2010,
  title = {Unified Picture for Magnetic Correlations in Iron-Based Superconductors},
  author = {Yin, Wei-Guo and Lee, Chi-Cheng and Ku, Wei},
  journal = {Phys. Rev. Lett.},
  volume = {105},
  issue = {10},
  pages  = {107004},
  numpages = {4},
  year = {2010},
  month = {Sep},
  publisher = {American Physical Society},
  doi = {10.1103/PhysRevLett.105.107004},
  url = {https://link.aps.org/doi/10.1103/PhysRevLett.105.107004}
}

@article{Dai2015,
  title = {Antiferromagnetic order and spin dynamics in iron-based superconductors},
  author = {Dai, Pengcheng},
  journal = {Rev. Mod. Phys.},
  volume = {87},
  issue = {3},
  pages  = {855--896},
  numpages = {42},
  year = {2015},
  month = {Aug},
  publisher = {American Physical Society},
  doi = {10.1103/RevModPhys.87.855},
  url = {https://link.aps.org/doi/10.1103/RevModPhys.87.855}
}

@article{Fernandes2022,
  title = {Iron pnictides and chalcogenides: a new paradigm for superconductivity},
  volume = {601},
  ISSN = {1476-4687},
  url = {http://dx.doi.org/10.1038/s41586-021-04073-2},
  DOI = {10.1038/s41586-021-04073-2},
  number  = {7891},
  journal = {Nature},
  publisher = {Springer Science and Business Media LLC},
  author = {Fernandes,  Rafael M. and Coldea,  Amalia I. and Ding,  Hong and Fisher,  Ian R. and Hirschfeld,  P. J. and Kotliar,  Gabriel},
  year = {2022},
  month = jan,
  pages  = {35--44},
}

@article{Bhmer2022,
  title = {Nematicity and nematic fluctuations in iron-based superconductors},
  volume = {18},
  ISSN = {1745-2481},
  url = {http://dx.doi.org/10.1038/s41567-022-01833-3},
  DOI = {10.1038/s41567-022-01833-3},
  number  = {12},
  journal = {Nature Physics},
  publisher = {Springer Science and Business Media LLC},
  author = {B\"{o}hmer,  Anna E. and Chu,  Jiun-Haw and Lederer,  Samuel and Yi,  Ming},
  year = {2022},
  month = dec,
  pages  = {1412--1419},
}

@article{Giles-Donovan2025,
  title = {First-order preemptive {Ising}-nematic transition in {K$_5$Fe$_4$Ag$_6$Te$_{10}$}},
  author = {Giles-Donovan, N. and Chen, Y. and Fukui, H. and Manjo, T. and Ishikawa, D. and Baron, A. Q. R. and Chi, S. and Zhong, H. and Cao, S. and Tang, Y. and Wang, Y. and Lu, X. and Song, Y. and Birgeneau, R. J.},
  journal = {Phys. Rev. B},
  volume = {111},
  issue = {22},
  pages  = {224103},
  numpages = {8},
  year = {2025},
  month = {Jun},
  publisher = {American Physical Society},
  doi = {10.1103/PhysRevB.111.224103},
  url = {https://link.aps.org/doi/10.1103/PhysRevB.111.224103}
}

@article{Cao2024,
  title = {Superstructures and magnetic order in heavily {Cu}-substituted {(Fe$_{1-x}$Cu$_x$)$_{1+y}$Te}},
  author = {Cao, Saizheng and Ma, Xin and Yuan, Dongsheng and Tao, Zhen and Chen, Xiang and He, Yu and Valdivia, Patrick N. and Wu, Shan and Su, Hang and Tian, Wei and Aczel, Adam A. and Liu, Yaohua and Wang, Xiaoping and Xu, Zhijun and Yuan, Huiqiu and Bourret-Courchesne, Edith and Cao, Chao and Lu, Xingye and Birgeneau, Robert and Song, Yu},
  journal = {Phys. Rev. B},
  volume = {109},
  issue = {4},
  pages  = {045142},
  numpages = {13},
  year = {2024},
  month = {Jan},
  publisher = {American Physical Society},
  doi = {10.1103/PhysRevB.109.045142},
  url = {https://link.aps.org/doi/10.1103/PhysRevB.109.045142}
}

@article{Lei2011,
  title = {Antiferromagnetism in semiconducting {KFe$_{0.85}$Ag$_{1.15}$Te$_2$} single crystals},
  author = {Lei, Hechang and Bozin, Emil S. and Wang, Kefeng and Petrovic, C.},
  year = {2011},
  month = aug,
  journal = {Phys. Rev. B},
  volume = {84},
  number  = {6},
  pages  = {060506},
  publisher = {American Physical Society},
  doi = {10.1103/PhysRevB.84.060506}
}

@article{Ang2013,
  title = {Electronic structure of the iron chalcogenide {KFeAgTe$_2$} revealed by angle-resolved photoemission spectroscopy},
  author = {Ang, R. and Nakayama, K. and Yin, W.-G. and Sato, T. and Lei, Hechang and Petrovic, C. and Takahashi, T.},
  year = {2013},
  month = oct,
  journal = {Phys. Rev. B},
  volume = {88},
  number  = {15},
  pages  = {155102},
  publisher = {American Physical Society},
  doi = {10.1103/PhysRevB.88.155102}
}

@article{Song2019,
  title = {Intertwined magnetic and nematic orders in semiconducting {KFe$_{0.8}$Ag$_{1.2}$Te$_2$}},
  author = {Song, Yu and Cao, Huibo and Chakoumakos, B. C. and Zhao, Yang and Wang, Aifeng and Lei, Hechang and Petrovic, C. and Birgeneau, Robert J.},
  year = {2019},
  month = feb,
  journal = {Phys. Rev. Lett.},
  volume = {122},
  number  = {8},
  pages  = {087201},
  publisher = {American Physical Society},
  doi = {10.1103/PhysRevLett.122.087201}
}

@article{Song2019a,
  title = {Strain-induced spin-nematic state and nematic susceptibility arising from {$2\times2$} {Fe} clusters in {KFe$_{0.8}$Ag$_{1.2}$Te$_2$}},
  author = {Song, Yu and Yuan, Dongsheng and Lu, Xingye and Xu, Zhijun and {Bourret-Courchesne}, Edith and Birgeneau, Robert J.},
  year = {2019},
  month = dec,
  journal = {Phys. Rev. Lett.},
  volume = {123},
  number  = {24},
  pages  = {247205},
  publisher = {American Physical Society},
  doi = {10.1103/PhysRevLett.123.247205}
}

@article{Bao2011,
  title = {A novel large moment antiferromagnetic order in {K$_{0.8}$Fe$_{1.6}$Se$_2$} superconductor},
  volume = {28},
  ISSN = {1741-3540},
  url = {http://dx.doi.org/10.1088/0256-307X/28/8/086104},
  DOI = {10.1088/0256-307x/28/8/086104},
  number  = {8},
  journal = {Chinese Physics Letters},
  publisher = {IOP Publishing},
  author = {Bao,  Wei and Huang,  Qing-Zhen and Chen,  Gen-Fu and Wang,  Du-Ming and He,  Jun-Bao and Qiu,  Yi-Ming},
  year = {2011},
  month = aug,
  pages  = {086104}
}

@article{Ye2011,
  title = {Common crystalline and magnetic structure of superconducting {A$_2$Fe$_4$Se$_5$} ({$A=$ K, Rb, Cs, Tl}) single crystals measured using neutron diffraction},
  author = {Ye, F. and Chi, S. and Bao, Wei and Wang, X. F. and Ying, J. J. and Chen, X. H. and Wang, H. D. and Dong, C. H. and Fang, Minghu},
  journal = {Phys. Rev. Lett.},
  volume = {107},
  issue = {13},
  pages  = {137003},
  numpages = {5},
  year = {2011},
  month = {Sep},
  publisher = {American Physical Society},
  doi = {10.1103/PhysRevLett.107.137003},
  url = {https://link.aps.org/doi/10.1103/PhysRevLett.107.137003}
}

@article{Wang2011,
  title = {Antiferromagnetic order and superlattice structure in nonsuperconducting and superconducting {Rb$_y$Fe$_{1.6+x}$Se$_2$}},
  author = {Wang, Meng and Wang, Miaoyin and Li, G. N. and Huang, Q. and Li, C. H. and Tan, G. T. and Zhang, C. L. and Cao, Huibo and Tian, Wei and Zhao, Yang and Chen, Y. C. and Lu, X. Y. and Sheng, Bin and Luo, H. Q. and Li, S. L. and Fang, M. H. and Zarestky, J. L. and Ratcliff, W. and Lumsden, M. D. and Lynn, J. W. and Dai, Pengcheng},
  year = {2011},
  month = sep,
  journal = {Phys. Rev. B},
  volume = {84},
  number  = {9},
  pages  = {094504},
  publisher = {American Physical Society},
  doi = {10.1103/PhysRevB.84.094504}
}

@article{mazin2010,
  title = {Superconductivity Gets an Iron Boost},
  author = {Mazin, Igor I.},
  year = {2010},
  month = mar,
  journal = {Nature},
  volume = {464},
  number  = {7286},
  pages  = {183--186},
  publisher = {Nature Publishing Group},
  issn = {1476-4687},
  doi = {10.1038/nature08914},
  urldate = {2025-06-25},
  copyright = {2010 Springer Nature Limited},
  langid = {english}
}

@article{si2016,
  title = {High-Temperature Superconductivity in Iron Pnictides and Chalcogenides},
  author = {Si, Qimiao and Yu, Rong and Abrahams, Elihu},
  year = {2016},
  month = mar,
  journal = {Nature Reviews Materials},
  volume = {1},
  number  = {4},
  pages  = {16017},
  publisher = {Nature Publishing Group},
  issn = {2058-8437},
  doi = {10.1038/natrevmats.2016.17},
  urldate = {2025-06-25},
  copyright = {2016 Macmillan Publishers Limited},
  langid = {english}
}

@article{Wang2014,
  title = {Two spatially separated phases in semiconducting {Rb$_{0.8}$Fe$_{1.5}$S$_2$}},
	volume = {90},
	ISSN = {1550-235X},
  url = {https://doi.org/10.1103/PhysRevB.90.125148},
  doi = {10.1103/PhysRevB.90.125148},
	number  = {12},
	journal = {Phys. Rev. B},
	publisher = {American Physical Society (APS)},
	author = {Wang, Meng and Tian, Wei and Valdivia, P. and Chi, Songxue and Bourret-Courchesne, E. and Dai, Pengcheng and Birgeneau, R. J.},
	year = {2014},
	month = sep,
  pages  = {125148},
}

@article{Bao2013,
  title = {Superconductivity tuned by the iron vacancy order in {K$_x$Fe$_{2-y}$Se$_2$}},
	volume = {30},
	ISSN = {1741-3540},
	url = {http://dx.doi.org/10.1088/0256-307X/30/2/027402},
	DOI = {10.1088/0256-307x/30/2/027402},
	number  = {2},
	journal = {Chinese Physics Letters},
	publisher = {IOP Publishing},
	author = {Bao, Wei and Li, Guan-Nan and Huang, Qing-Zhen and Chen, Gen-Fu and He, Jun-Bao and Wang, Du-Ming and Green, M. A. and Qiu, Yi-Ming and Luo, Jian-Lin and Wu, Mei-Mei},
	year = {2013},
	month = feb,
	pages  = {027402}
}

@article{Kazakov2011,
  title = {Uniform patterns of {Fe}-vacancy ordering in the {K$_x$(Fe,Co)$_{2-y}$Se$_2$} superconductors},
  volume = {23},
  ISSN = {1520-5002},
  url = {http://dx.doi.org/10.1021/cm201203h},
  DOI = {10.1021/cm201203h},
  number  = {19},
  journal = {Chemistry of Materials},
  publisher = {American Chemical Society (ACS)},
  author = {Kazakov, Sergey M. and Abakumov, Artem M. and Gonz{\'a}lez, Santiago and Perez-Mato, Juan Manuel and Ovchinnikov, Alexander V. and Roslova, Maria V. and Boltalin, Alexander I. and Morozov, Igor V. and Antipov, Evgeny V. and Van Tendeloo, Gustaaf},
  year = {2011},
  month = sep,
  pages  = {4311--4316},
}

@article{Wang2016_2,
	title = {Elucidating the magnetic and superconducting phases in the alkali metal intercalated iron chalcogenides},
	author = {Wang, Meng and Yi, Ming and Tian, Wei and Bourret-Courchesne, Edith and Birgeneau, Robert J.},
	journal = {Phys. Rev. B},
	volume = {93},
	issue = {7},
	pages  = {075155},
	numpages = {8},
	year = {2016},
	month = {Feb},
	publisher = {American Physical Society},
	doi = {10.1103/PhysRevB.93.075155},
	url = {https://link.aps.org/doi/10.1103/PhysRevB.93.075155}
}

@article{Tam2020,
  title = {Orbital selective spin waves in detwinned {NaFeAs}},
  author = {Tam, David W. and Yin, Zhiping and Xie, Yaofeng and Wang, Weiyi and Stone, M. B. and Adroja, D. T. and Walker, H. C. and Yi, Ming and Dai, Pengcheng},
  journal = {Phys. Rev. B},
  volume = {102},
  issue = {5},
  pages  = {054430},
  numpages = {11},
  year = {2020},
  month = {Aug},
  publisher = {American Physical Society},
  doi = {10.1103/PhysRevB.102.054430},
  url = {https://link.aps.org/doi/10.1103/PhysRevB.102.054430}
}

\end{document}